**A Multi-Agent Framework Integrating Large Language Models and Generative AI for Accelerated Metamaterial Design**

Jie Tian[1], Martin Taylor Sobczak[1], Dhanush Patil[1], Jixin Hou[1], Lin Pang[1], Arunachalam Ramanathan[1], Libin Yang[1], Xianyan Chen[2], Yuval Golan[3], Xiaoming Zhai[4], Hongyue Sun[1], Kenan Song[1*], Xianqiao Wang[1*]

[1.] School of ECAM, University of Georgia, Athens, GA 30602, USA

[2.] Department of Biostatistics and Epidemiology, University of Georgia, Athens, GA 30602, USA

[3.] Department of Materials Engineering, and the Ilse Katz Institute for Nanoscale Science and Technology, Ben-Gurion University of the Negev, Beer-Sheva 8410501, Israel

[4.] Department of Mathematics, Science, and Social Studies Education, College of Education, University of Georgia, Athens, GA 30622

Corresponding authors: kenan.song@uga.edu; xqwang@uga.edu

## Abstract:

Metamaterials, renowned for their exceptional mechanical, electromagnetic, and thermal properties, hold transformative potential across diverse applications, yet their design remains constrained by labor-intensive trial-and-error methods and limited data interoperability. Here, we introduce CrossMatAgent—a novel multi-agent framework that synergistically integrates large language models with state-of-the-art generative AI to revolutionize metamaterial design. By orchestrating a hierarchical team of agents—each specializing in tasks such as pattern analysis, architectural synthesis, prompt engineering, and supervisory feedback—our system leverages the multimodal reasoning of GPT-4o alongside the generative precision of DALL-E 3 and a fine-tuned Stable Diffusion Extra Large (XL) model. This integrated approach automates data augmentation, enhances design fidelity, and produces simulation- and 3D printing-ready metamaterial patterns. Comprehensive evaluations, including Contrastive Language–Image Pre-training (CLIP)-based alignment, SHAP (SHapley Additive exPlanations) interpretability analyses, and mechanical simulations under varied load conditions, demonstrate the framework's ability to generate diverse, reproducible, and application-ready designs. CrossMatAgent thus establishes a scalable, AI-driven paradigm that bridges the gap between conceptual innovation and practical realization, paving the way for accelerated metamaterial development.

# 1 Introduction

Metamaterials have garnered attention in recent years due to their extraordinary properties across various fields[1–4]. These materials exhibit exceptional properties, including distinctive electromagnetic[5–8], thermal[9], and mechanical[10–12] properties, presenting substantial promise to overcoming the limitations of traditional materials. Moreover, traditional design methods are still labor-intensive, relying heavily on trial-and-error processes and hindered by insufficient data interoperability. The advent of machine learning (ML) and large language models (LLMs) offers new opportunities for metamaterial design, which can explore the target design space with the highest possibility of designated performance. However, despite significant efforts and the accumulation of knowledge, there is still no research indicating that we can apply the scaling law[11,13], where LLM performance benefits from the scale increase of dataset and model size. This mainly arises from the fact that protein design has a cohesive representation that enables research findings from different domains to enhance ML development, whereas metamaterials do not.

In comparison with traditional methods in metamaterial design, ML enables researchers to explore vast and complex design spaces that are otherwise infeasible with conventional methods. For example, at the earlier stage of ML employment, algorithms such as regression [6], genetic algorithms (GA)[14–16], and artificial neural networks (ANN)[17] were used to optimize the design parameters of metamaterial structures to achieve targeted properties, but the target design space (e.g., parametric design for truss-based metamaterials) is relatively simple. Though design space is limited, one branch of ANN - physics-informed neural networks (PINN)[17] enhances the intuitive understanding of the relationship between input variation and properties of designed metamaterial with a carefully crafted loss function. To enhance the design space for metamaterial, later, the utilization of algorithms such as convolutional neural networks (CNNs) and generative adversarial networks (GANs), demonstrate superior performance in image processing, forward performance prediction[18], and model inverse design[18–20]. Particularly, there is a substantial enhancement in the realm of inverse design, for example, facilitating pixel-wise design capabilities. More recent ML models like variational autoencoders (VAEs)[21–23] offer an innovative approach to discovering new metamaterial structures and optimizing current designs by overcoming the design space limitation further, which makes the discrete design space continuous in latent space. Overall, these approaches allow designers to accurately predict material behaviors, automate repetitive tasks, and reduce reliance on labor-intensive trial-and-error methods when designing and discovering new metamaterials. However, the issue remains that despite the extensive effort invested and the vast amount of data and knowledge accumulated, there is still no indication of a groundbreaking algorithm, such as the generative pretrained transformer (GPT)[24–26], that can harness the scaling law and utilize this accumulated knowledge across domains and algorithms for further discoveries. Inspired by the semantic design in a recent work the most intuitive representation of metamaterials is a plain text description; this format is domain-agnostic, unrestricted by design space, and inherently compatible with LLMs. In this instance, a meticulously crafted descriptive schema in human language has the potential to connect the previously isolated domains of metamaterials, thereby enabling them to benefit from advancements in LLMs.

A common concern in the current machine learning-empowered study of metamaterials is the perception that all small domains in a metamaterial structure function as isolated islands[1,27–29]. For each particular design and application, unique algorithms and representations are formulated, especially for some salient aspects of metamaterial design – nature-inspired designs[30,31]. As a result, the broader potential of ML in unifying and streamlining metamaterial design processes remains underutilized. Examining the bioinformatics field[32,33], which benefits from the surge of large language models (LLMs). The Simplified Molecular Input Line Entry System (SMILES)[11] allows molecular chains to be depicted as sequences of characters, making them suitable for LLMs naturally. Motivated by advancements in multimodal LLMs[26,31,31,34,35], including vision transformers (ViT)[34], particularly the outstanding capabilities of innovative commercial models like GPT-4o[36], Claude-3.5-sonnet, and Gemini-1.5-pro, state-of-the-art LLMs exhibit advanced performance in reasoning about the physical world[37], mechanics[36], etc. Employing state-of-the-art LLMs can help us construct a concise and clear description of metamaterials despite the domain or application. In conclusion, LLMs possess promising potential in bridging isolated domains and fostering cross-domain development knowledge[24,38].

A general framework for LLMs, specifically, a method that constructs interaction graphs from complex input-output data, known as a multi-agent framework, while demonstrating high in-context learning capacity[31,36], has shown great capability in scientific analysis. Popular tools like Auto-Gen[39], Langchain[40], and Llama-index[41] all demonstrate outstanding performance. The process of fine-tuning[42–46] pre-trained models represents an instrumental method for utilizing the knowledge embedded within LLMs. Even a limited amount of labeled data for fine-tuning can substantially enhance the performance of general LLMs in specialized domains without the need to train a new model from scratch. Effectively, the fine-tuning process alters the weights of a select number of Multi-Layer Perceptron (MLP) layers at the model's head or, alternatively, incorporates additional MLP[43] layers that can be readily activated or deactivated. In particular, X-LoRA[42], developed by Buehler, integrates multiple LoRA[47] (Low-Rank Adaptation) that have been trained on diverse datasets, resulting in distinguished performance. In conclusion, advanced models such as GPT-4o possess the capability to comprehend images, texts, and even physical laws. This technology represents a unified solution for navigating the domains of metamaterial design, with tools developed to leverage the optimal advantages of artificial intelligence. However, there are three main unsolved difficulties. The first is demonstrating an understanding of metamaterials, as all current text encoders for large language models (LLMs) are trained using text datasets such as Wikipedia. It is imperative to ensure that the models account for the geometry of metamaterials rather than relying solely on context. Secondly, this involves the challenge of acquiring a high-quality dataset that encompasses various types of metamaterials. The final challenge pertains to application. Recent research indicates that, although image generation models such as Dall-E can provide promising images of models, they remain conceptual demonstrations and cannot be utilized for model building, simulation, or manufacturing[11,48].

In this paper, we present CrossMatAgent, an innovative multi-agent framework designed to enhance the metamaterial design process by leveraging the multimodal capabilities of LLMs. The challenges are addressed through a structured three-step framework design process. Initially, by establishing a collaborative multi-agent system, we capitalize on the advancements in image generation technology, particularly Dall-E 3. This approach employs data collected from existing research to facilitate the automatic generation of datasets. The reasoning and reflection mechanisms inherent in LLMs assist in creating paired data that establish a correlation between the generated images and their corresponding geometrical and morphological significances. Subsequently, a fine-tuning methodology enhances the machine ML model’s intrinsic understanding of geometrical and morphological information, thereby eliminating the necessity for step-by-step reasoning. Ultimately, by utilizing the datasets generated through the multi-agent framework and fine-tuning a foundational model—Stable Diffusion XL (SDXL)—we achieve the successful production of designs that are nearly prepared for computational simulation and 3D printing. Furthermore, the fine-tuned SDXL model exhibits exceptional performance in few-shot learning when integrated with LoRA, particularly with a limited dataset, where the style of metamaterials can be effectively altered. Finally, we integrate the fine-tuned model with GPT-4o to form CrossMatAgent as a cohesive agent. This integrated system empowers users to design metamaterials based on image or instruction inputs and demonstrates remarkable performance in zero-shot learning.

# 2 Methodology

## 2.1 Development of the CrossMatAgent Framework

As the cornerstone of this manuscript, vision language models (VLMs), such as GPT-4o, along with their advanced reasoning capabilities, empower LLMs to comprehend fundamental mechanics involved in engineering applications and facilitate the design of metamaterials. Our CrossMatAgent framework is built upon the state-of-the-art general-purpose multimodal LLM, GPT-4o, through the OpenAI API. The development of CrossMatAgent involves three vital steps. We start with a straightforward yet effective traditional team hierarchy rooted in GPT-4o and Dall-E 3, to evaluate and illustrate the performance of a multi-agent system (MAS), benefiting from reflection and ongoing dialogue[49,50]. As illustrated in Figure 1a, the team has developed a focus on analyzing the geometric and morphological properties of the pattern derived from the specified image, which was collected from other research publications regarding metamaterials, lattice structures[51,52], and nature-inspired designs[29,53], among others. The format of messages exchanged between agents, along with the overall cost and the total number of chat messages, is fixed. The final goal is to generate a sharp and clear image pattern that requires only a few simple steps of computer vision (CV) to transition into a printable model. Aside from that, within the team, agents are specifically designed to play a pivotal role in addressing the comprehensive task, which encompasses the automation of generation and revision. Consequently, it can generate a collection of readily usable images of metamaterial patterns that possess required geometrical and morphological properties. This step can be considered a data augmentation process for collecting metamaterial patterns from various formats and domains.

The second step utilizes the augmented dataset to fine-tune a pre-trained Stable Diffusion XL (SDXL) model, as depicted in Figure 1b. While Dall-E 3 exhibits superior performance in general image generation, it lacks the specificity required for metamaterial designs suitable for computational simulation and/or 3D printing. Fine-tuning SDXL on the augmented dataset enhances its ability to produce outputs that meet specific design objectives. By aligning textual prompts with visual patterns during training, the model adapts to generate highly targeted designs that fulfill the requirements for manufacturability. This tailored approach not only replaces the reliance on Dall-E 3 but also ensures that the outputs are more directly usable for practical applications.

In the final step, we develop CrossMatAgent, which can utilize the optimal features of the refined SDXL model to generate new metamaterial designs. In the most basic two-agent team, a ChatGPT agent functions as a prompt interpreter and enhancer, working alongside CrossMatAgent, as depicted in Figure 1c, to validate the performance of the designed agent and prepare for more complex MAS with human proxy as input.

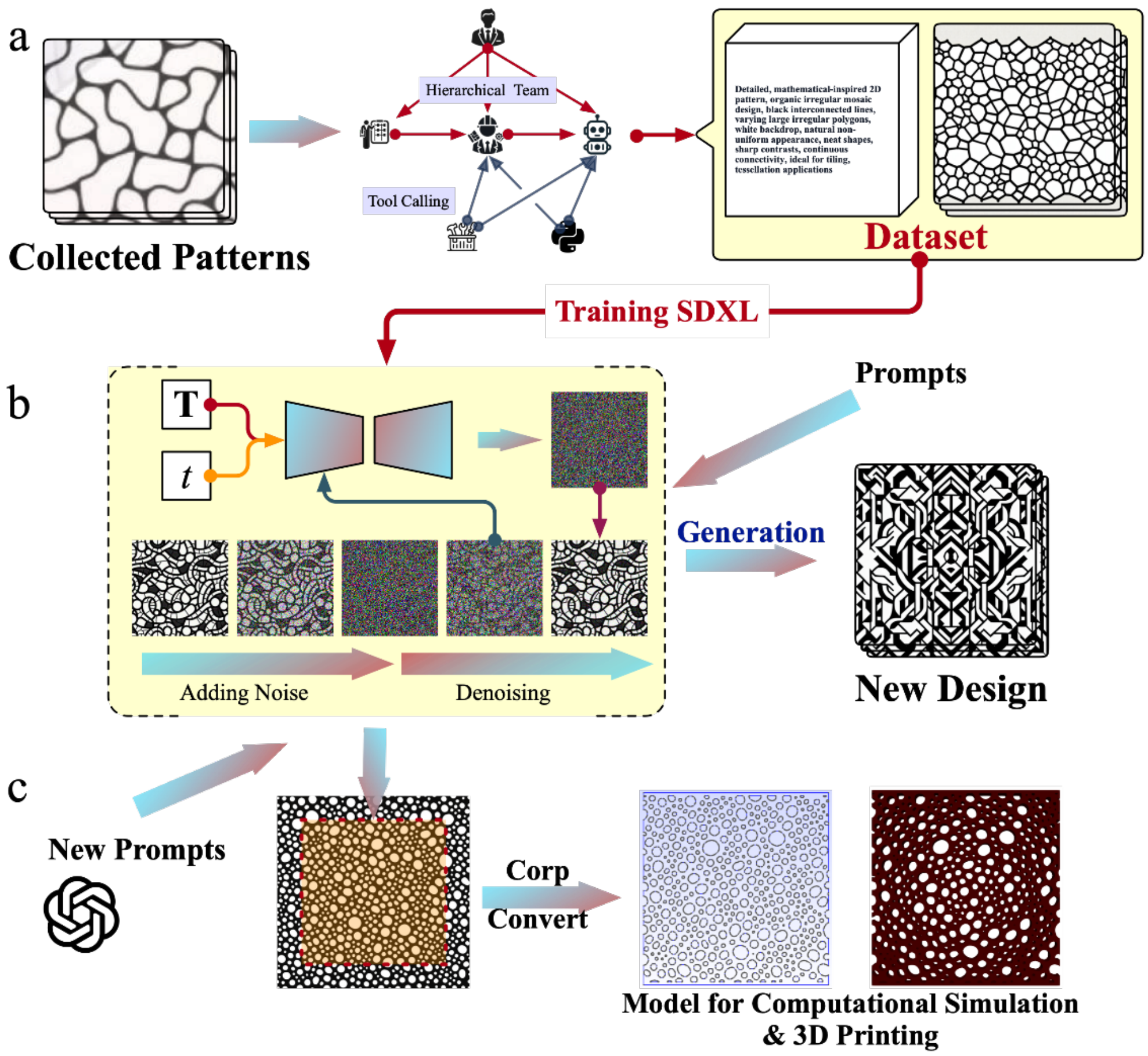

**Figure 1: Overview of the CrossMatAgent framework.** (a) Leveraging a collection of diverse patterns (additional examples provided in the Table S3), the multi-agent system framework generates new, similar patterns based on corresponding text prompts, functioning as a data augmentation. (b) Utilizing the augmented data, we fine-tune a diffusion model which builds on the pre-trained Stable Diffusion XL model from Stability AI[54]. (c) In the final stage, the fine-tuned model is employed to generate new designs, with ChatGPT serving as an interpreter and prompt refiner.

### *2.1.1. Agent Roles and Collaborative Workflow in the Multi-Agent System*

As shown in Figure 2, we build an agent-based team comprised of four primary agents—**Describer**, **Architect**, **Builder**, and **Supervisor** —working collaboratively to analyze input patterns, generate prompts, and produce outputs tailored for Dall-E 3. Each agent's role and responsibilities are further detailed in Table S1, where system prompts guide their specific contributions to the workflow. We use LangChain[55,56] to build the MAS which is well known for the advantages of structure design and communications between agents and well support for various LLMs. In LangChain, an agent typically consists of three components: LLM backend, system prompt, and tools for calling. Here, we use OpenAI GPT-4o as a backend for all agents, ensuring high-performance reasoning and seamless workflow execution.

The **Describer Agent** initiates the process by thoroughly analyzing the input pattern. This involves a general assessment, geometric segmentation (e.g., identifying repeating units), and morphological evaluation under various mechanical conditions such as tension or compression (as indicated in Table S2), as detailed in Table S1. These insights are conveyed to the **Architect Agent**, who utilizes expertise in mechanics and structural design to interpret the Describer's analysis. The Architect concentrates on generating a comprehensive textual description of the geometry, connectivity, and functional aspects of the pattern, ensuring the structural integrity and practical relevance of the design.

The **Builder Agent**, a specialist in prompt engineering, utilizes the Architect's detailed descriptions to create prompts optimized for Dall-E 3. This agent focuses on generating visually clear and mathematically coherent 2D patterns that highlight key structural elements such as connectivity and symmetry. By concentrating on patterns with specific characteristics like bio-inspired connectivity or mechanical stability, the Builder ensures that the outputs meet practical requirements for manufacturability, simulation, and/or 3D printing. Finally, the **Supervisor Agent** oversees the entire workflow, reviewing the outputs for quality and alignment with the input objectives. It examines all the chat history, including generated images. If discrepancies arise, the Supervisor offers iterative feedback by providing specific instructions to one agent in the loop, enabling the loop to jump to that point and continue refining the design to ensure that the generated patterns adhere to all specifications. Since the Supervisor reads all the chat history, it is less likely that a MAS will meet the requirements when the chat history is extensive. Therefore, we manually set the upper bound of chat history to 20.

This team's hierarchical design, as illustrated in Figure 2, allows the MAS to systematically process input patterns, generate diverse yet consistent variations, and ensure that the outputs are suitable for further applications. The structured use of prompts and specialized roles, detailed in Table S1, enables the system to tackle complex design challenges, expanding the potential for innovative metamaterial development while maintaining practicality and precision. It is important to note that the system prompts for agents are designed within the GPTs tool in ChatGPT, which provides an interactive approach to agent design.

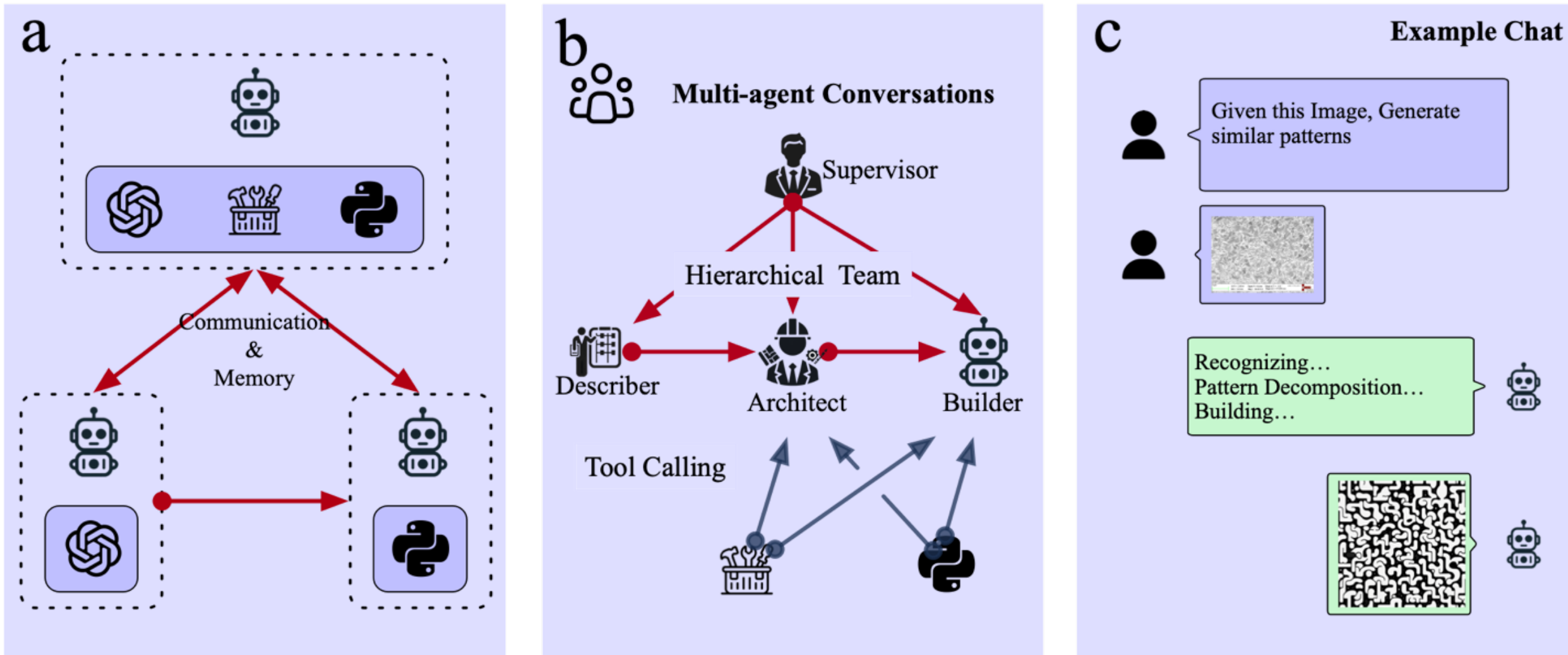

**Figure 2: Agent roles and collaborative workflow in the multi-agent system.** (a) In a multi-agent system, in order to maintain output stability, a supervisory role is required, and the work is divided into four parts: (i) the Describer analyzes the input structure's pattern and provides a description; (ii) the Architect constructs the model based on the Describer's description; (iii) the Builder interprets the Architect's design and converts it into a prompt for Dall-E-3; (iv) the Supervisor oversees the entire process to ensure that the results align with the input properties; if discrepancies are found, the Supervisor suggests a restart point and provides detailed instructions. Details of the prompt are presented in Table S1. (b) The multi-agent design consists of two primary components: the first involves the goals and tools specific to each agent, where each agent is equipped with a designated purpose and the necessary tools to fulfill it; the second encompasses the structural and communication workflow within the team. (c) The usage diagram illustrates the functions and communication flow of the packaged system.

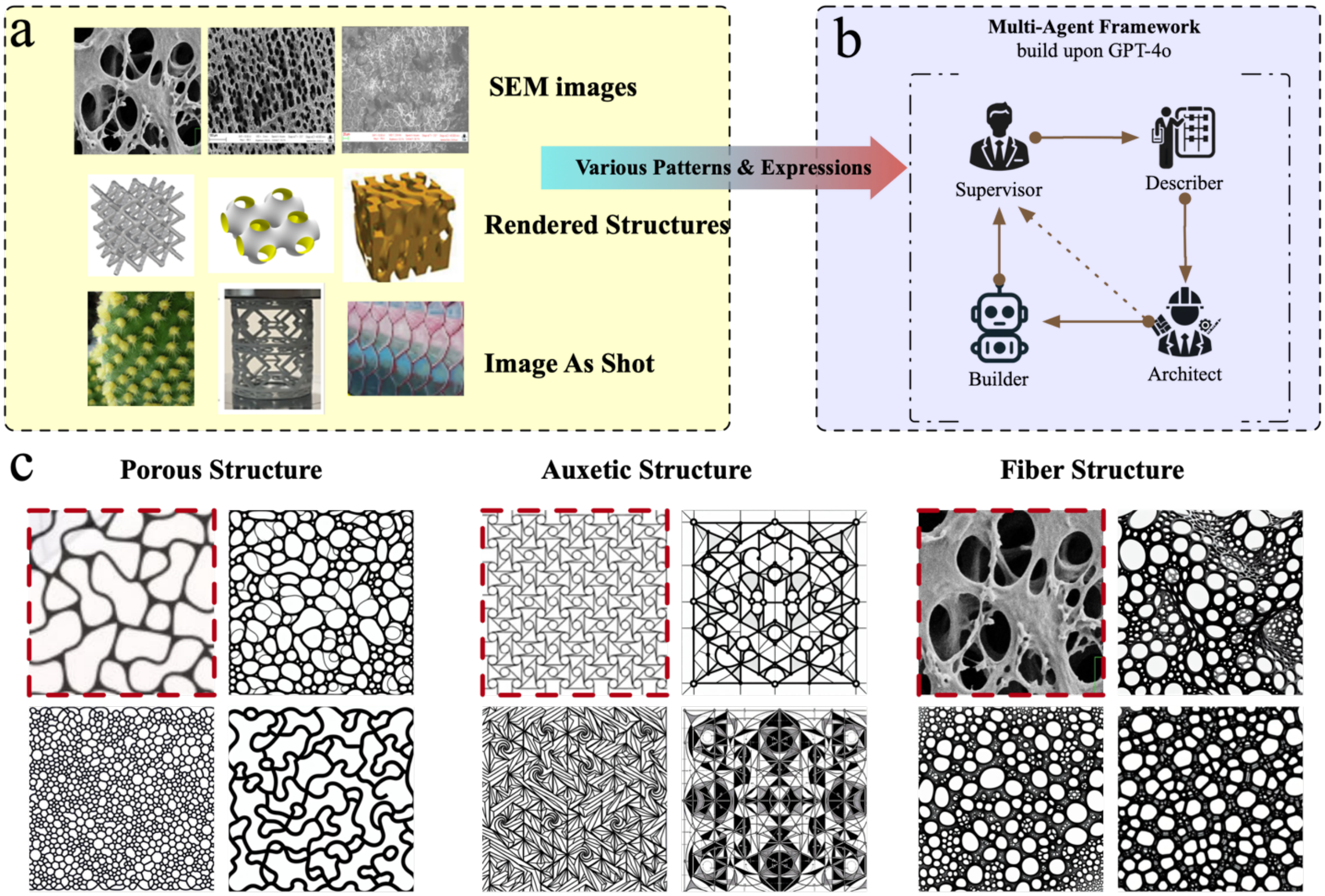

**Figure 3: Robust input processing in the multi-agent system.** (a) During data collection from existing research, various patterns of metamaterials are identified. This paper primarily focuses on three types of metamaterials: SEM images, 3D/2D rendered model images, and shot images. Special emphasis is placed on SEM images as they are grayscale, aligning with the characteristics of the model we aim to generate. (b) In the collection process, the types and structures of metamaterials are not controlled, and a MAS is employed to produce a unified representation of these materials, followed by generating additional patterns that share similar features. (c) The various inputs for the Measurement and Analysis System (MAS) are presented within the red rectangle, and the corresponding outputs from the MAS demonstrate its capabilities.

*2.1.2. Self-Correction and Robust Input Processing in the Multi-Agent System*

As detailed in Table S2, a MAS conversation commences with an image input, subsequently followed by the Describer's geometric and morphological analysis. Thereafter, the Architect presents a potential structural description for the model. The Builder then constructs the prompt for DALL-E 3, which generates a new pattern. Ultimately, the Supervisor reviews the workflow and identifies the restart point, providing comments if necessary; otherwise, the workflow concludes. This design of MAS allows agents to perform self-correction based on the results of image generation. Moreover, it designates which team members should take responsibility and offers further instructions and guidance for the next steps from that juncture. Owing to the self-correction capabilities and reasoning abilities of GPT-4o, the MAS effectively extracts the geometrical and morphological properties from diverse input formats. This robustness allows for seamless information retrieval across various data sources. As illustrated in Figure 3a, the system processes a wide range of inputs, including scanning electron microscopy (SEM) images, three-dimensional (3D) and two-dimensional (2D) rendered models, as well as photographic representations of metamaterials obtained from several review articles[27,57–59]. These inputs undergo comprehensive analysis to identify key characteristics such as geometry, texture, and connectivity within the MAS framework (Figure 3b). Notably, SEM images are prioritized due to their grayscale nature, which aligns well with the generative capabilities of Dall-E 3 and the emphasis on 2D structural representation.

As illustrated in Figure 3c, various input images, which represent porous, auxetic, and fiber structures, along with their corresponding results from MAS, indicate that the MAS possesses the capability to produce a diverse array of design variations that retain the essential characteristics of the input data in conjunction with the respective text prompt descriptions. The MAS ensures that the variations retain essential morphological characteristics, such as the web-like connectivity of bio-inspired patterns or the intricate void structures typical of porous materials. By producing consistent yet diverse designs, the framework expands the design space while maintaining fidelity to the defining properties of the original input. This capability not only facilitates the exploration of new metamaterial patterns but also bridges disparate research results, providing a unified approach to understanding and innovating within the field. The resulting dataset comprises approximately 700 image-text pairs; the example and dataset address are shown in Table S3. However, the dataset size is inherently constrained by the iterative nature of prompt engineering, requiring multiple generation cycles to refine usable output – thereby increasing API-related computational costs.

## 2.2 High-Fidelity Metamaterials Synthesis

### *2.2.1 CLIP-Enhanced Text-to-Image Alignment for Geometrical Design Understanding*

Our next step involves generating novel metamaterial designs that are closely aligned with the MAS system's generation process. The CLIP[60] (Contrastive Language-Image Pre-Training) model offers an innovative mechanism to align feature distributions between text prompts and images, facilitating the creation of new image designs by sampling from the textual design space. This outlines the architecture of the CLIP model, with its dual-encoder setup enabling the text and image modalities to map into a shared latent space.

In the context of stable diffusion models like SDXL, illustrated in Figure 1b, the UNet processes two primary inputs: an encoded text prompt and a random noise vector, generating a final output that represents the most probable result sampled from the design space corresponding to the text prompt. In other words, the text encoder of CLIP is a fundamental element for understanding the design space when leveraging SDXL in the creation of metamaterials. To train the CLIP model, we first need to transfer the text descriptions from the Dall-E 3 version into split phrase segments that better suit SDXL based on GPT-4 (Table S3). We use "OpenAI/CLIP-ViT-Large-Patch14", a popular CLIP base model, as the base model for fine-tuning. The total training step of fine-tuning is 400, and the similarity matrix between prompts (shown in Figure 4b) indicates a good training. As depicted in Figure 5a, a well-trained CLIP model should identify the most suitable phrase from a given set to describe key features of a metamaterial. Conversely, it should also identify the metamaterial from a group that best matches the given description.

### *2.2.2. UNet Architecture and Training Optimization for High Quality Metamaterial Generation*

At the heart of the **SDXL** model lies the UNet model, which serves as the core generative engine for producing high-quality metamaterial designs. The UNet in the diffusion models is trained to reverse a stochastic forward noising process to gradually restore a data point $X_0$ (image of metamaterial, can you point it out in the Figures), from the latent data distribution $X_0 \sim q(X)$ to a prior distribution, usually a standard Gaussian distribution.

In the training process, techniques such as gradient clipping, adaptive learning rate scheduling, and data augmentation stabilize the process and prevent overfitting. In the data augmentation process, both random flipping and random cropping are employed in response to the constraints posed by the limited training data. These methods enhance the UNet's robustness, allowing it to generalize effectively across diverse metamaterial patterns.

### 2.3 Design and Application of CrossMatAgent

The utilization of a proficiently trained SDXL model, which is informed by descriptions and designs from diverse domains and originates from various forms, is crucial for leveraging the extensive knowledge it has acquired. Aided by tools for constructing microservices such as FastAPI[61], we successfully encapsulated the SDXL as a service similar to Dall-E. By utilizing the compilation of prompts from the dataset as a reference library for GPT-4o to enhance the prompts for SDXL, and subsequently employing SDXL as a tool for generation, we develop the **CrossMatAgent**. Since all the prompts for SDXL are generated by GPT-4o, the prompt library serves more as guidance rather than a query library. As illustrated in Figure 6, we present two methods for deploying CrossMatAgent: the simple two-agent system shown in Figure 6a and the complex multi-agent system derived from the MAS, depicted in Figure 6b.

In a straightforward two-agent application (Figure 6a), a manager is assigned to elucidate the human request, which subsequently leads to the formation of **CrossMatAgent**. Furthermore, as illustrated in Figure 6b, the intricate framework, which encompasses the primary architecture of the Multi-Agent System (MAS), incorporates two principal features: human proxy and dynamic graph, aimed at enhancing adaptivity and versatility. The human proxy enables researchers to incorporate additional information after the supervisor's revision, while the dynamic graph facilitates more intricate input. For instance, when only textual descriptions are provided, the system has the capability to bypass the Describer step.

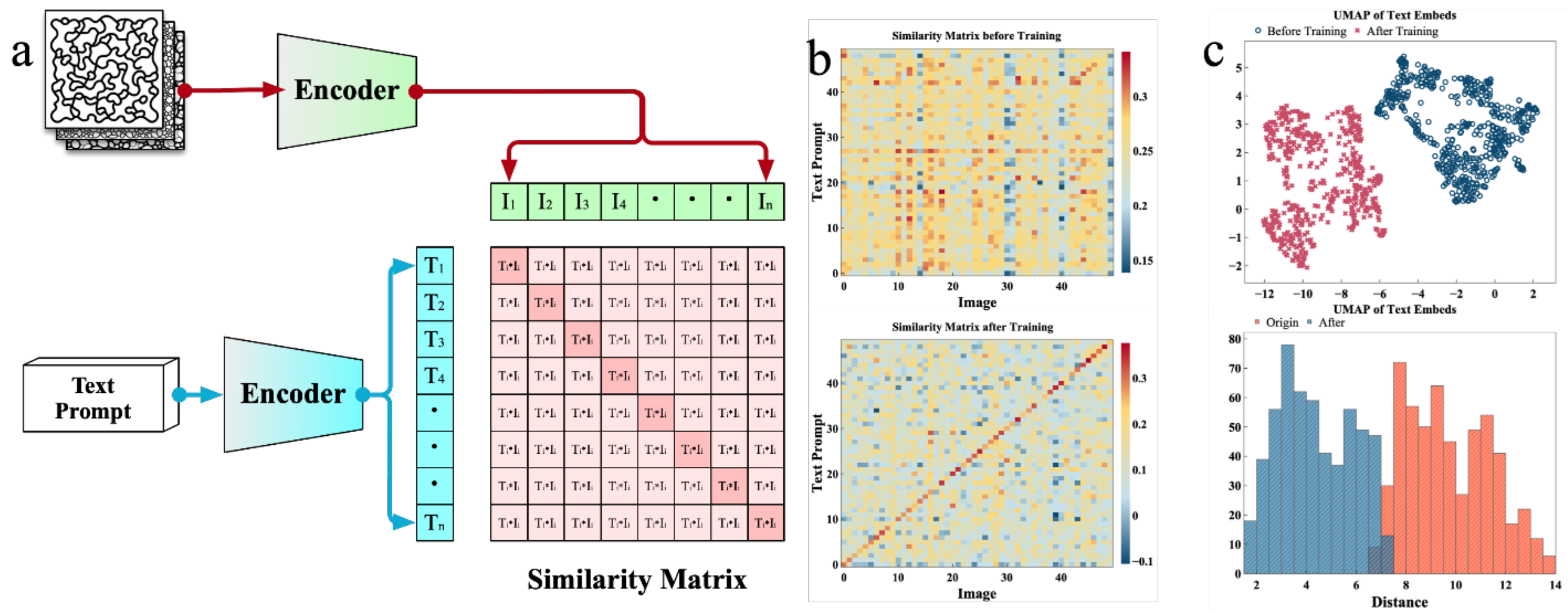

**Figure 4: Architecture of CLIP model and training results.** (a) OpenAI originally proposed the Contrastive Language-Image Pre-Training (CLIP) model, which uses two pre-trained encoders—one for text and one for images. After the encoding process, two sequences of features are obtained to represent the image and text, followed using two MLP blocks to align these features and optimize the KL divergence. (b) The similarity matrix between the prompt phrases and the corresponding images is presented before and after the training process. (c) The distribution of prompts within latent space is presented both with and without training.

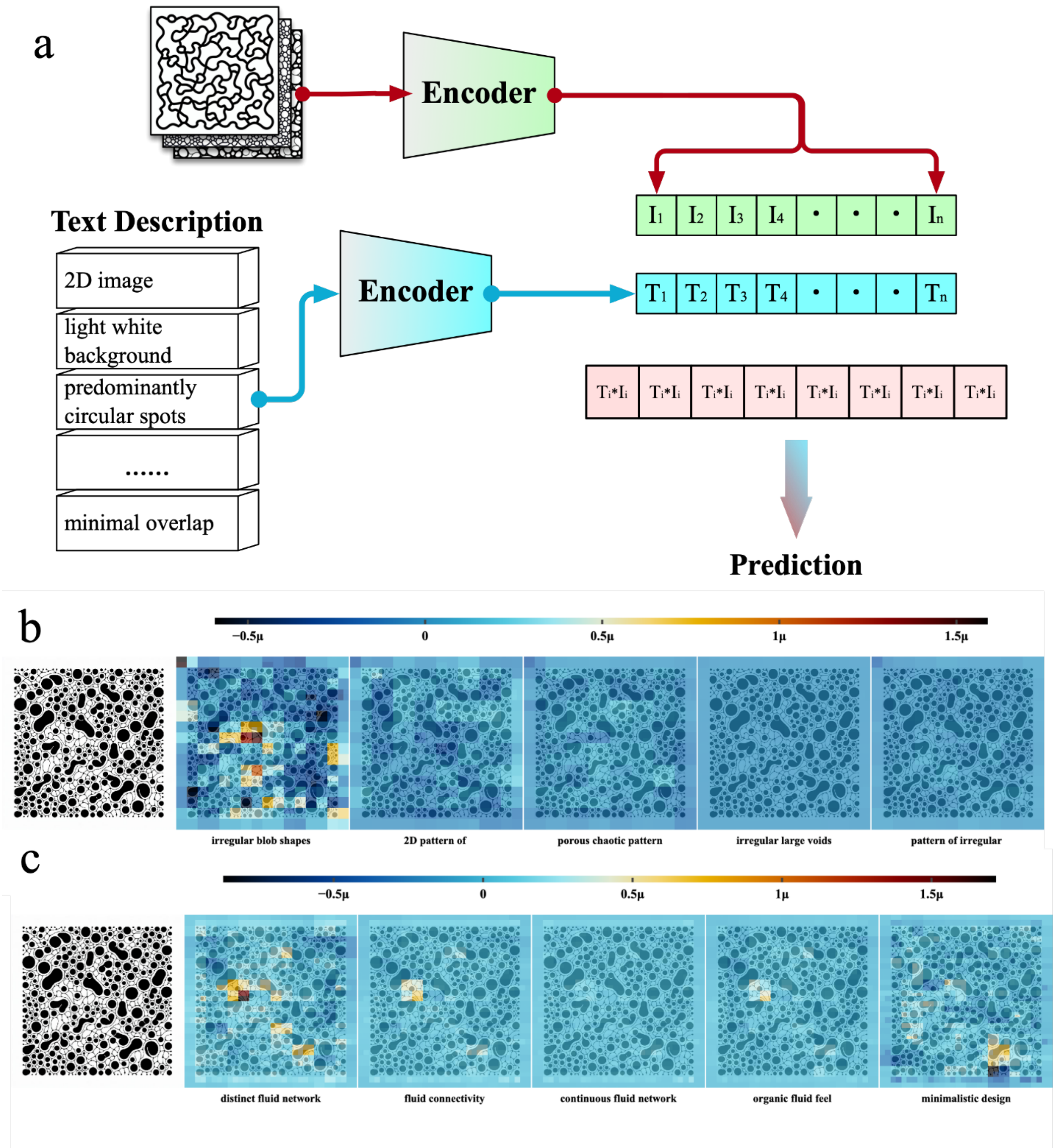

**Figure 5: Application of the CLIP model and its understanding of patterns.** (a) The diagram illustrates the prediction process, where the CLIP model uses a similarity matrix to identify the most accurate description. (b, c) SHAP analysis reveals the most significant features and the regions of the image responsible for these features. We analyze both OpenAI's original CLIP model (b) and a fine-tuned version (c), alongside the corresponding ground truth text prompts. The fine-tuning process enhances the model's ability to understand patterns. PS:2D pattern, irregular, organic shapes with curved edges, blobs, randomly distributed, interconnected via thin lines, fluid, network-like structure, black shapes and lines, white background, stark contrast, continuous connectivity, web-like appearance, nature-inspired patterns

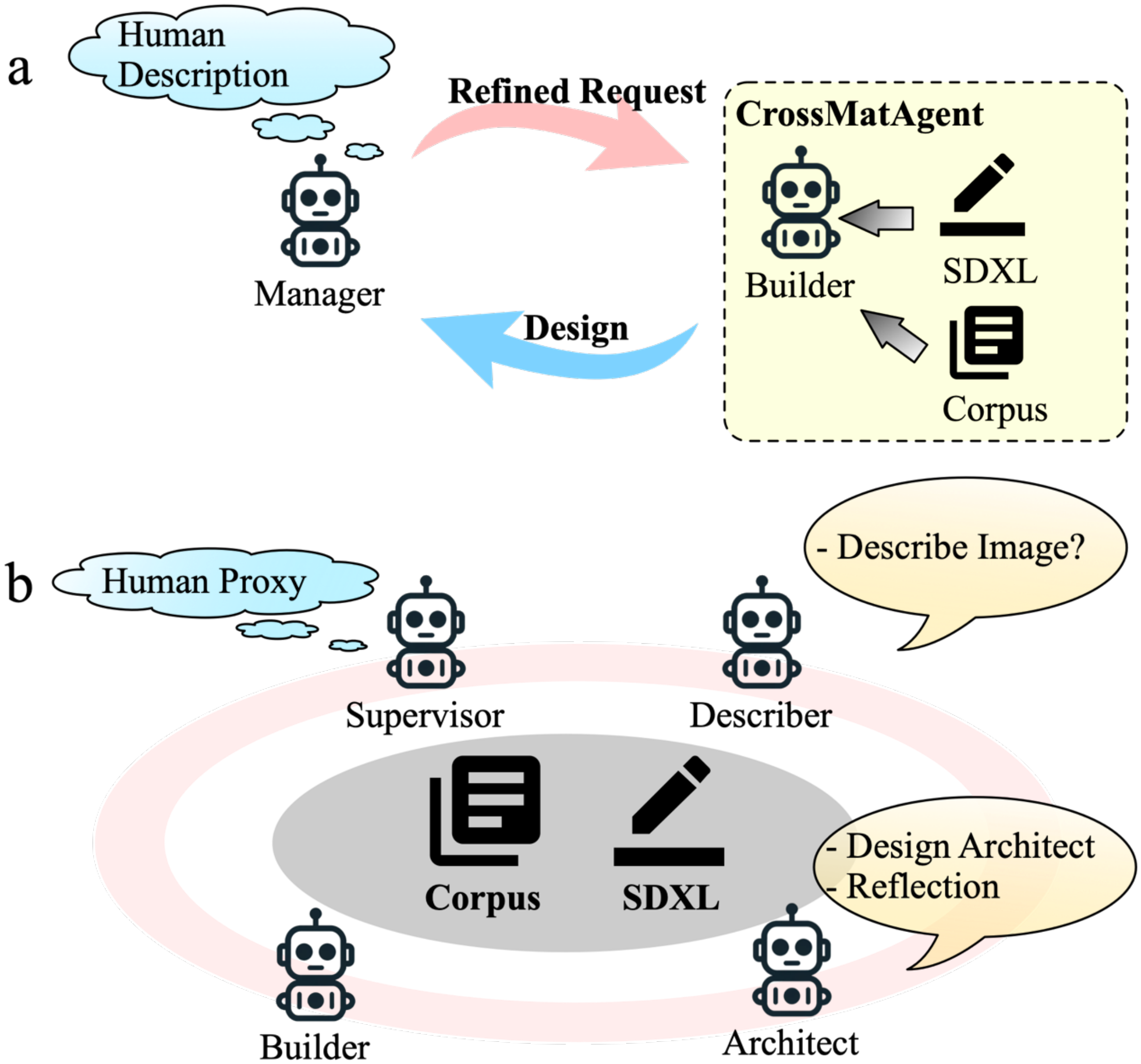

**Figure 6: Design and application of CrossMatAgent.** (a) The SDXL is presented as a tool for developers, accompanied by an initial process to create an SDXL form prompt. The most straightforward two-agent conversational system, featuring a human description, involves the primary managerial agent processing the information and formulating a more precise and coherent request tailored for the builder (CrossMatAgent). (b) The improved architecture of the Multi-Agent System (MAS) includes a human proxy, which enables users to furnish supplementary information throughout the iterative process. This enhancement offers each agent a broader spectrum of options in their actions, while still relying on the CrossMatAgent framework.

# 3 Results & Discussions

## 3.1 CLIP Training Results

### *3.1.1 Semantic Analysis of Model's Understanding in Prompt*

To examine how the text encoder comprehends the text description both before and after training, we employ two complementary approaches. First, we assess the alignment between the text space and the metamaterial design space. Second, we evaluate whether the CLIP model is capable of accurately matching the design with the prompt from a collection of mixed pairs. Figure 4b shows the similarity matrix between the text prompts and the designs, indicating an advanced performance improvement from training. To assert whether the remarkable performance arises from overfitting, we conduct an in-depth analysis of the embedding results obtained from the text encoder. As illustrated in Figure 4c, we visualize the distribution of all prompts across the entire dataset using UMAP (Uniform Manifold Approximation and Projection for Dimension Reduction). While the distribution of distances between data points, as illustrated in the lower panel of Figure 4c, does not exhibit significant differences before and after training, a more distinct signature of clustering pattern emerges post-training, signifying improved organization within the embedding space.

Further to understand the improvement after training, we investigate the distribution of latent space. To have a uniform latent vector size, we apply a standardized padding scheme using "<|endoftext|>" to ensure the length of all tokenized phrases is 17. Then, we employ the UMAP method to project the latent vector into two-dimension, followed by the application of K-Means clustering, which has $K = 8$ for better presentation, as shown in Figure 7a and Figure 7b. The distinction derived from the K-Means clustering is minimal, which is intuitively understandable, considering that the original CLIP model is inherently state-of-the-art in natural language processing (NLP). Thus, we perform Principal Component Analysis (PCA) to identify the principal axes of each cluster. Subsequently, we adhere to the PCA values to aggregate representative phrases within each cluster (shown in Figure 7c and 7d). It must be acknowledged that, during this process, the phrases with the highest PCA values in each cluster exhibit considerable similarity in both meaning and vocabulary; therefore, we manually select representations after the processing. From the representative phrases, it is evident that the foundation model already possesses a significant understanding of geometrical concepts; for example, it can cluster phrases like "puzzle piece" and "scientific diagram" together. Here, additional training could further improve the understanding in metamaterial; for instance, a trained version of CLIP categorizes "interconnected" and "overlap" as distinct types, whereas the base model classifies "overlap" and "2D" together. Interestingly, it is observed that, within the base model, the length of a phrase has a significant influence other than its meaning; however, this influence has been diminished in the training process. Notably, in extreme cases, such as excessively long texts, this length continues to exert a more pronounced effect. This phenomenon matches the outer layer illustrated in Figures 7a and 7b.

In conclusion, the training process of the CLIP model is pivotal for the comprehension of geometrical and morphological information, and it significantly contributes to the generation of high-quality material that aligns with the intentions of the researcher.

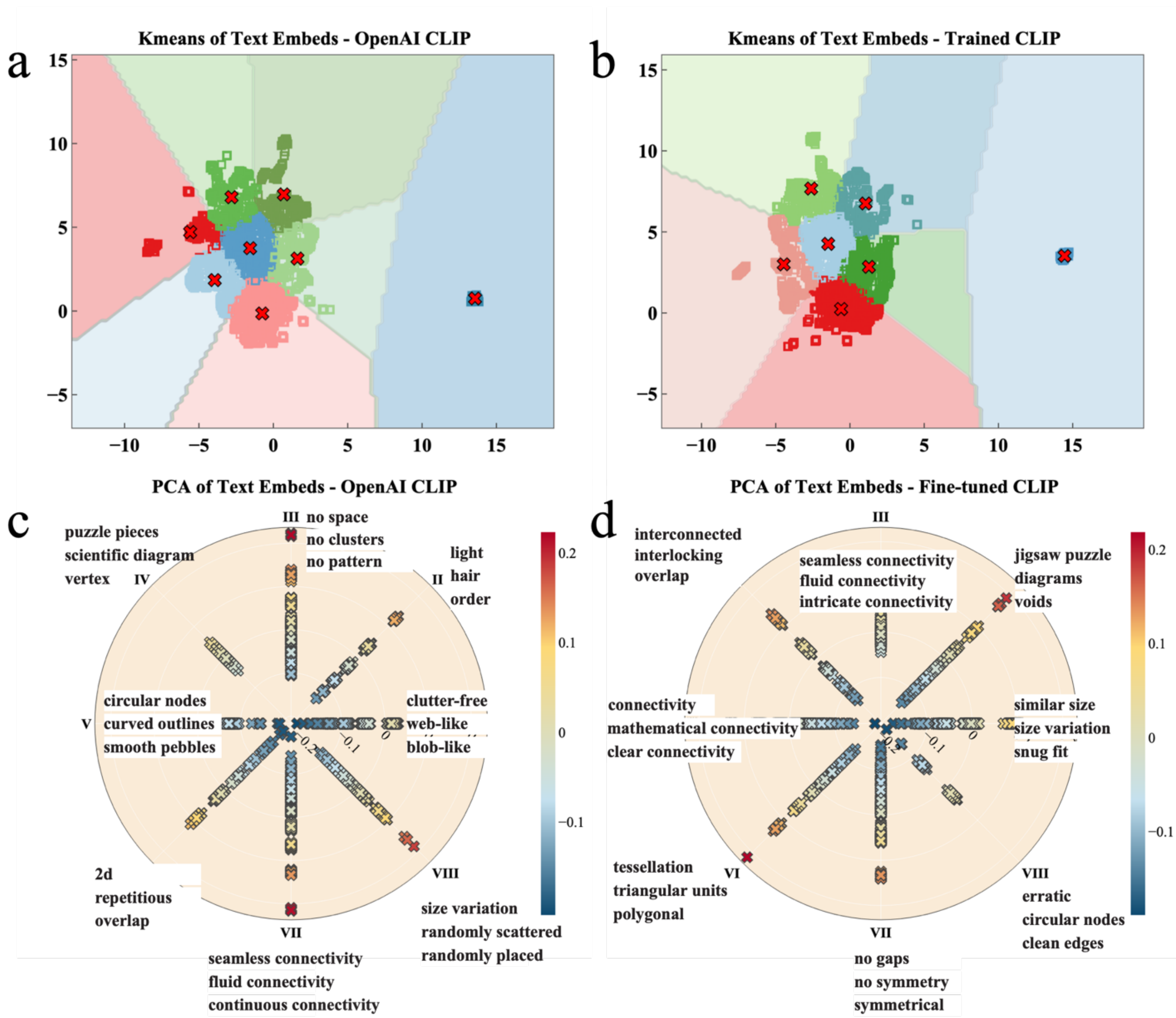

**Figure 7: Latent space analysis of text prompts.** To gain a deeper understanding of how the ML model interprets pattern descriptions, we extract the latent features from the text encoder in the CLIP model. UMAP is then used to project these features into two dimensions. After the projection, K-Means analysis is applied to divide the data into 8

clusters (a, b). For each cluster, PCA analysis is performed to identify the most representative phrase of each cluster (c, d).

### *3.1.2 Assessment of CLIP's Text-to-Image Alignment via SHAP Analysis*

To further evaluate the CLIP model's ability to interpret designs' features and connect them with textual descriptions, we perform SHAP (SHapley Additive exPlanations) analysis. Through the examination of the gradient variation that occurs when utilizing a square mask on the design image, it is possible to analyze the different regions of the design and identify those that are most influential in relation to specific textual prompts. This approach provides a comprehensive analysis of how the model establishes its text-image correlations. In this context, we emphasize the distribution of SHAP values pertaining to the CLIP model rather than their absolute quantities. For instance, if elevated values are observed consistently across the image, it indicates that the overarching pattern is predominant and suggests a greater robustness of the CLIP model. As illustrated in Figure 5b, c, the comparison of the SHAP analysis results indicates a considerable enhancement in the understanding of geometrical patterns. This is particularly evident when juxtaposing the most appropriate phrase selected by CLIP with the ground truth prompt, wherein the trained CLIP demonstrates a markedly superior performance. It is imperative to note that in constructing the evaluation, the CLIP model is employed to identify the most relevant phrase description rather than utilizing the entire prompt from the comprehensive dataset collection, with the aim of validating robustness.

A comprehensive analysis of the trained CLIP model is illustrated in Figure 8. We employ SHAP analysis utilizing a variety of mask sizes, specifically 16, 32, 64, and 128 from Figure 8a to 8d to assess the consistency of feature attributions. The prediction of best match phrases is almost the same as the varying mask sizes, which demonstrate the robustness of the trained CLIP model. Furthermore, the values of the masks are relatively homogeneous, indicating that the features contributing to the predictions are derived from the entirety of the design image rather than from a small segment. SHAP analysis elucidates the interpretative strength of the trained CLIP. By systematically deconstructing the contribution of each region within the image to the overall alignment, we discern a distinct emphasis on morphological details that are directly correlated with the descriptive phrases. This observation accentuates the efficacy of the fine-tuning process in augmenting the CLIP model's capability to interpret and align domain-specific data. Moreover, it demonstrates how the CLIP model fortifies the SDXL diffusion model by ensuring that the generative process remains anchored in the input descriptions, thereby enhancing the quality and relevance of the outputs.

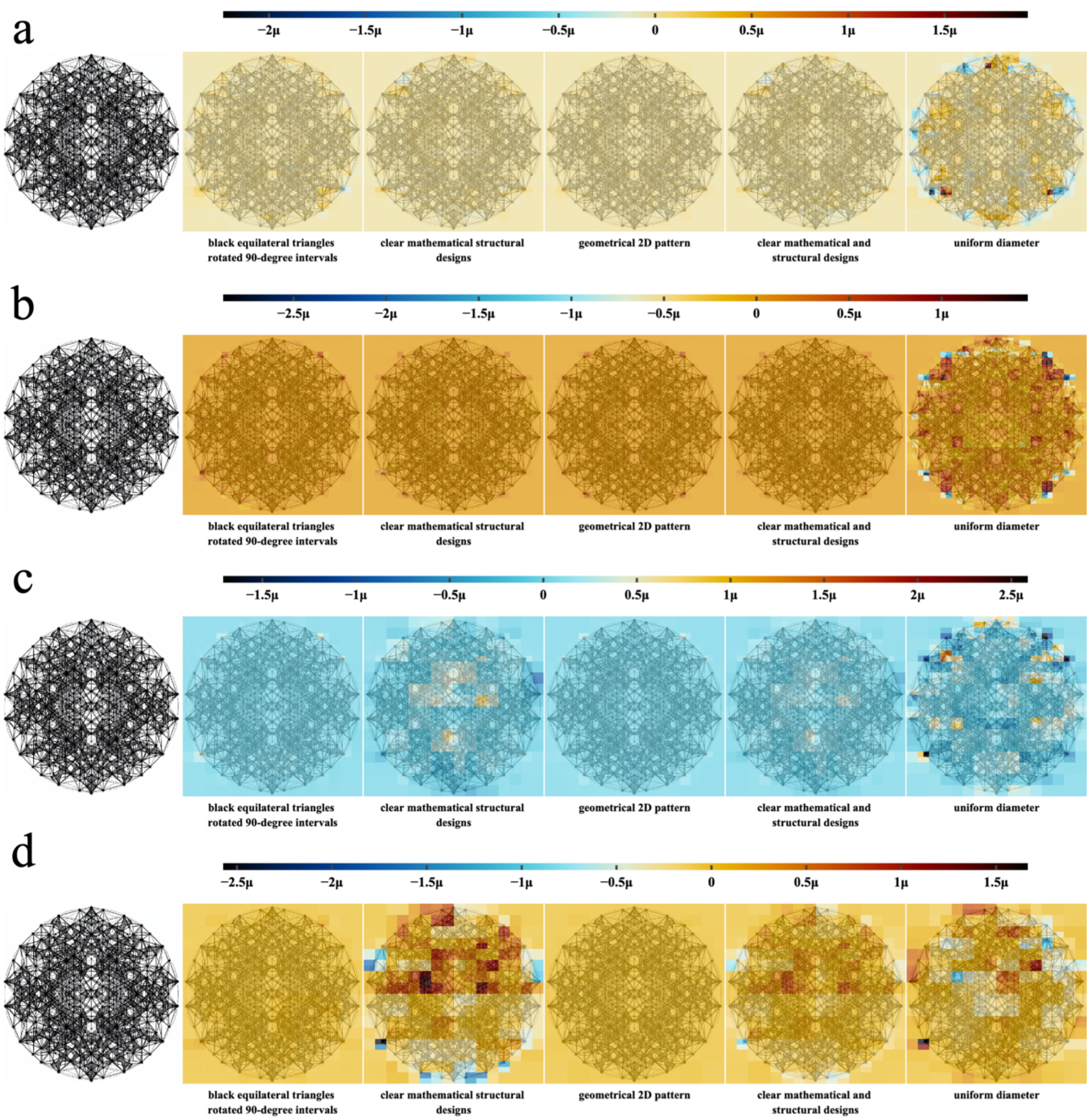

**Figure 8: SHAP analysis with different sizes of blur masks.** To further explore how the AI model interprets image and text features, we conduct SHAP analysis using various blur mask sizes ($16 \times 16$ to $128 \times 128$ ). Despite the differences in mask sizes, the primary text features identified remain consistent, demonstrating that the model can comprehend features across the entire image.

## 3.2 Metamaterial Design

### *3.2.1 Comparative Evaluation of CrossMatAgent with Different Generative Methods*

The performance of the CrossMatAgent, leveraging the fine-tuned SDXL model, is benchmarked against other generative approaches in Figure 9, illustrating the efficacy of the proposed methodology. The figure highlights the outputs from different stages of the workflow, including those generated by the multi-agent framework and the fine-tuned SDXL model, alongside traditional generative approaches. This comparison demonstrates the clear advantages of combining agent-driven data processing with fine-tuned generative AI models for metamaterial design. From Figures 9a through 9f, we present examples derived from the input source generated by ChatGPT within the webpage chat interface. The designs sourced from MAS merit attention, particularly as the three generations originate from a single iteration loop generated by the SDXL model. Additionally, these designs incorporate elements from CrossMatAgent and a continuous slice of several TPMS (Triply Periodic Minimal Surface) solid models exhibiting a sine wave surface.

In Figure 9a, we present three identical inputs for MAS; the right panel depicts the input corresponding to the results of the methods shown in Figures 9b-e. Given the significant variances in brightness, clarity, and resolution of the inputs, MAS displays advanced adaptability. The results shown in the ChatGPT chatbox (Figure 9b) illustrate the current challenges faced by LLMs; when given a pattern to mimic and produce, the generation still lacks manufacturability. Nevertheless, Dall-E 3 remains the state-of-the-art model, exhibiting significantly better performance compared to the SDXL base model (Figure 9d). While the designs are noteworthy and inspiring, holding significant promise, there is still an absence of a clear pathway from the generated output to a functional model. Also, the interactive chatting process for fine-tuning generation in chatbox is often uncontrollable and arduous. By leveraging the capabilities of multi-agents, the design of MAS (Figure 9c) reaches new heights, achieving the next generation of artificial intelligence performance, which can generate sharp, clear, and highly manufacturable designs. However, one problem remains: the ability to reproduce is still absent, and each generation with the same prompt will not yield the same results. The CrossMatAgent demonstrates a comprehensive understanding of geometric principles as they pertain to metamaterials, along with an exceptionally specialized capacity for designing these materials. As illustrated in Figure 9e, using the same prompt with different random seeds yields three results that show remarkable similarity. The resulting designs demonstrate stability and clarity due to training for metamaterial design, and importantly, they are reproducible, benefiting from the open-source model SDXL. The clarity of the produced images is sufficient for modeling in computational simulations and 3D printing. The designs presented in Figure 9f are indeed interchangeable; the relationship between the model and the design is strongly interlinked and can be articulated mathematically. This mathematical articulation serves as a distinct advantage over all machine learning models, not limited to just LLMs.

In summary, the comparison in Figure 9 underscores the importance of integrating agent-driven workflows with fine-tuned generative models. The multi-agent framework provides a strong foundation for data preparation and iterative refinement, while the fine-tuned SDXL model excels in generating high-quality, application-ready designs. This synergy between human-like interpretation and AI-driven generation represents a significant advancement in the field of metamaterial design, offering a scalable and efficient approach to bridging the gap between conceptualization and realization.

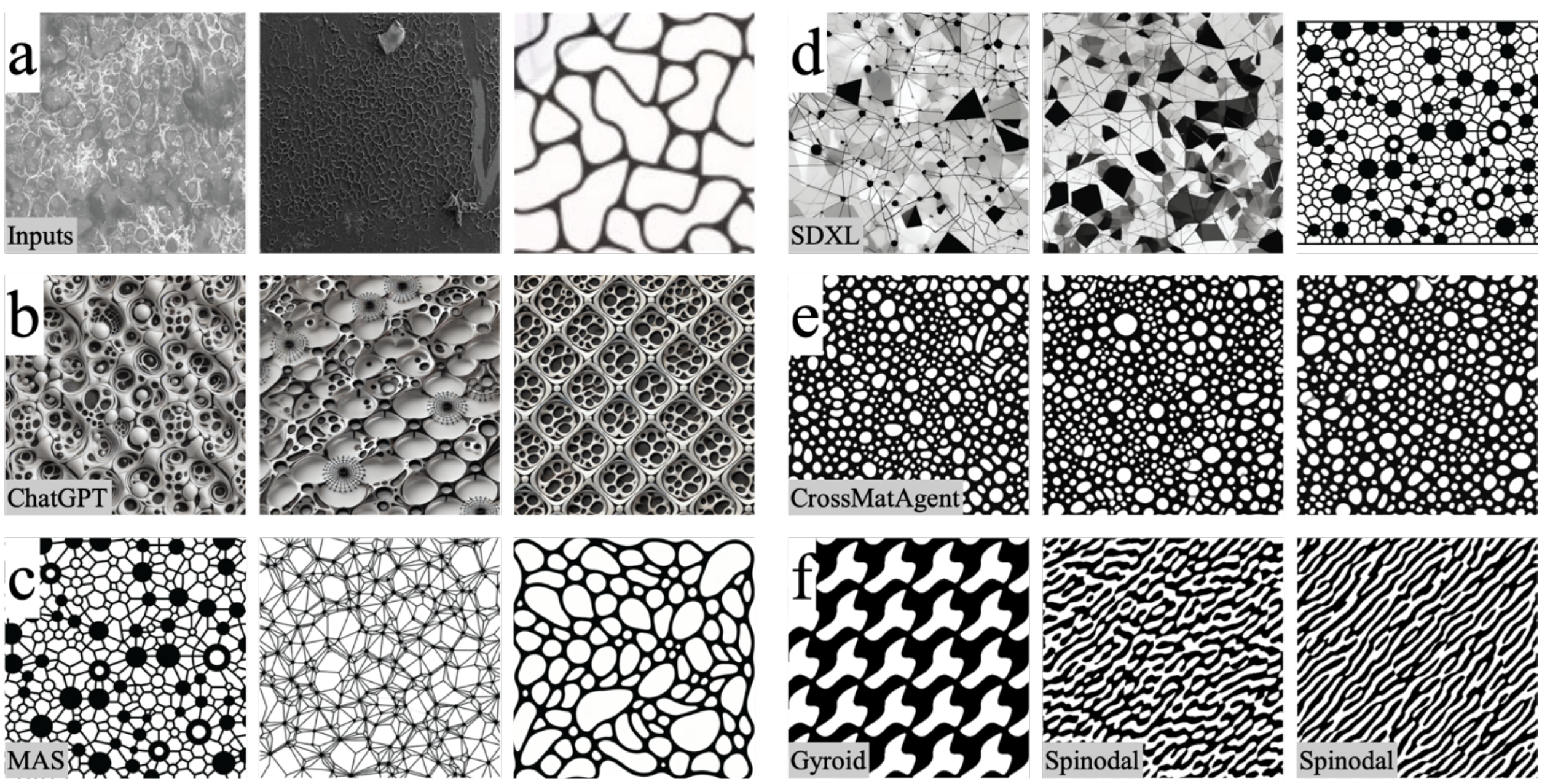

**Figure 9: Comparison between generations from different models and traditional methods.** (a) The collected image patterns, specifically SEM images, can be broadly categorized into bio-structures and porous structures. (b) With the advancements in LLMs, they can now accurately interpret image content. Here, we use ChatGPT to analyze the structure of input images and generate similar patterns; the detailed prompt is provided in Table S4. (c) The generation process using the multi-agent system requires three iterations to achieve the result, with the full process detailed in Figure 6. (d) When the model is not fine-tuned, the generation meets the expectations for metamaterial design. (e) The generated metamaterial designs are clear, crisp, and stable, and most importantly, they match the style of the input. (f) The spinodal structure is a well-known bio-inspired structure, recognized for its unique properties and prevalence in natural forms. A common method to derive a 2D pattern from spinodal structures involves using a curved plane to slice the 3D structure while ensuring the resulting pattern remains path connected.

### *3.2.2 Simulation Demonstration of Metamaterial Designs from CrossMatAgent*

One of the key contributions of this research is to provide a user-friendly and cohesive presentation of metamaterial design, thereby facilitating the development of various metamaterial designs from a single model and addressing the limitations present in LLMs concerning model creation. To further validate the utility of CrossMatAgent, we build a few final models and run some simple simulations using TPU 95A, a widely used 3D printing material with well-characterized mechanical properties.

As illustrated in Figure 10, we examine three representative metamaterial designs: a bio-inspired structure, a grid-like porous structure, and an auxetic structure. The building process is illustrated in Figure 1c, which comprises only two steps. The first step involves center-cropping to mitigate potential issues, followed by the application of edge polygon detection to identify the shape and location of all porous; in this way, we can avoid the non-path-connected model. This methodology enables the creation of structurally sound designs suitable for computational analysis. Three fundamental mechanical simulations are conducted: simple tension (first column), compression (second column), and shear (third column). The results pertaining to stress distribution are illustrated in Figure 10. These findings align with the expected mechanical response, with no errors observed during the meshing and other preprocessing phases of the simulation. It is noteworthy that the auxetic model exhibits a negative Poisson's ratio under compression, which constitutes a primary area of interest in the study of auxetic metamaterials. Also, we 3D printed the model with TPU 95A as a demonstration.

While CrossMatAgent is not yet a fully end-to-end surrogate modeling tool, this study demonstrates its capability to generate diverse metamaterial designs informed by research across multiple scientific domains. This validation underscores its potential as a powerful AI-driven tool for advancing metamaterial research and design automation.

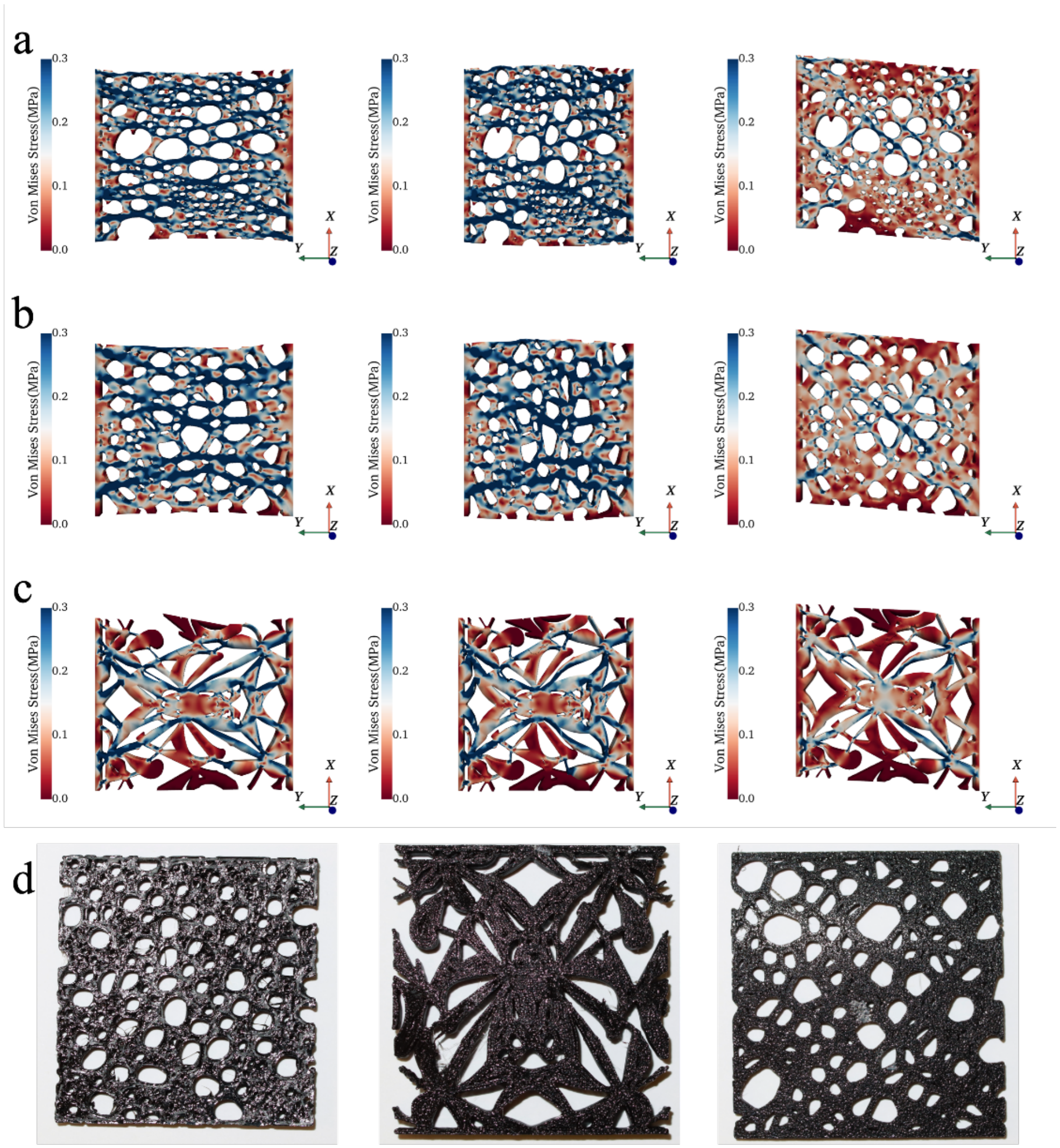

**Figure 10: Computational simulations of porous, bio-inspired, and auxetic structures generated by a diffusion model under compression, tension, shear loads and 3D-print demonstration.** (a) A typical porous structure is generated by adding random polygons using traditional methods. (b) A bio-inspired structure mimics cellular architecture. (c) An auxetic structure under compression exhibits rotation of certain components and buckling of connections, indicative of a negative Poisson's ratio or reduced stiffness. (d) The 3D-printed models, which serve as representations matching the simulation models, indicate the potential for fabrication.

### 3.3 Few-Shot Learning and Model Extensibility

One of the very important features that measure the robustness of an AI model is the performance of few-shot learning. To assess the extendibility of our approach, we evaluate performance under few-shot learning conditions using LoRA[42,43] as the fine-tuning method. As shown in Figure 11a, we select a set of waterlily images as a validation test, featuring a hierarchical linkage structure with multiple symmetry axes. The first step involves translating the authentic image of the waterlily into a form suitable for few-shot learning. We employ the MAS to produce a limited number of designs, as illustrated in Figure 11b, which contains merely 12 pairs of images and prompts. We then fine-tune the "waterlily" LoRA for 5000 steps on a single H100 GPU, which takes about 1 hour of training.

During the inference phase, the fine-tuned LoRA model is integrated with a weighting factor of 0.7. Designs are subsequently generated in a simple multi-agent context, as depicted in Figure 6a, with the final outputs presented in Figure 11c. The results demonstrate that the fine-tuned base model effectively captures the essential morphological characteristics of metamaterials. However, the MAS still exhibits some inherited limitations from DALL-E 3, affecting its precision in certain cases (Figure 11b). Notably, SDXL exhibits strong performance in few-shot learning and possesses the unique capability to combine multiple LoRAs, enabling the generation of multi-property fusion designs. This property significantly enhances the model's extensibility, paving the way for more versatile and adaptable metamaterial generation.

Furthermore, we evaluate the CrossMatAgent within the framework of the multi-agent system depicted in Figure 6b, utilizing the previously established LoRA configuration for the waterlily. The system is tasked with generating metamaterials that resemble "waterlily-like" characteristics, and the outcomes are illustrated in Figure 12. The first row (Figure 12a-c) displays the generation, while the second row (Figure 12d-f) presents the results after center cropping and image processing, including binarization and noise removal. The final 3D printing results are shown in Figure 12g-i. In summary, the performance of CrossMatAgent is remarkably impressive, particularly in light of the similarity observed before and after image processing. This observation signifies its high capability and manufacturability.

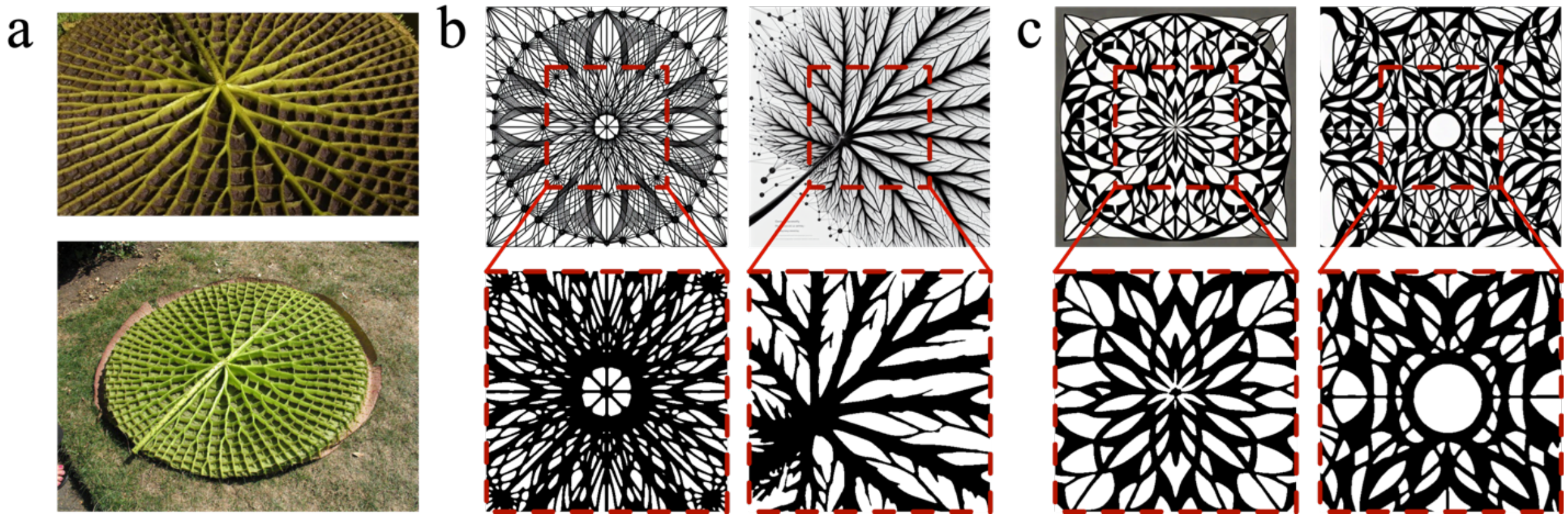

**Figure 11: Model extensibility tested on waterlily structure**. (a) In the provided example of a water lily image, we can anticipate a hierarchical structure and a link-based framework essential for an effective design. (b) The findings derived from MAS are relatively complex due to Dall-E 3. The designs generated from LoRA when applied to CrossMatAgent are clear and sharp, effectively capturing the key features.

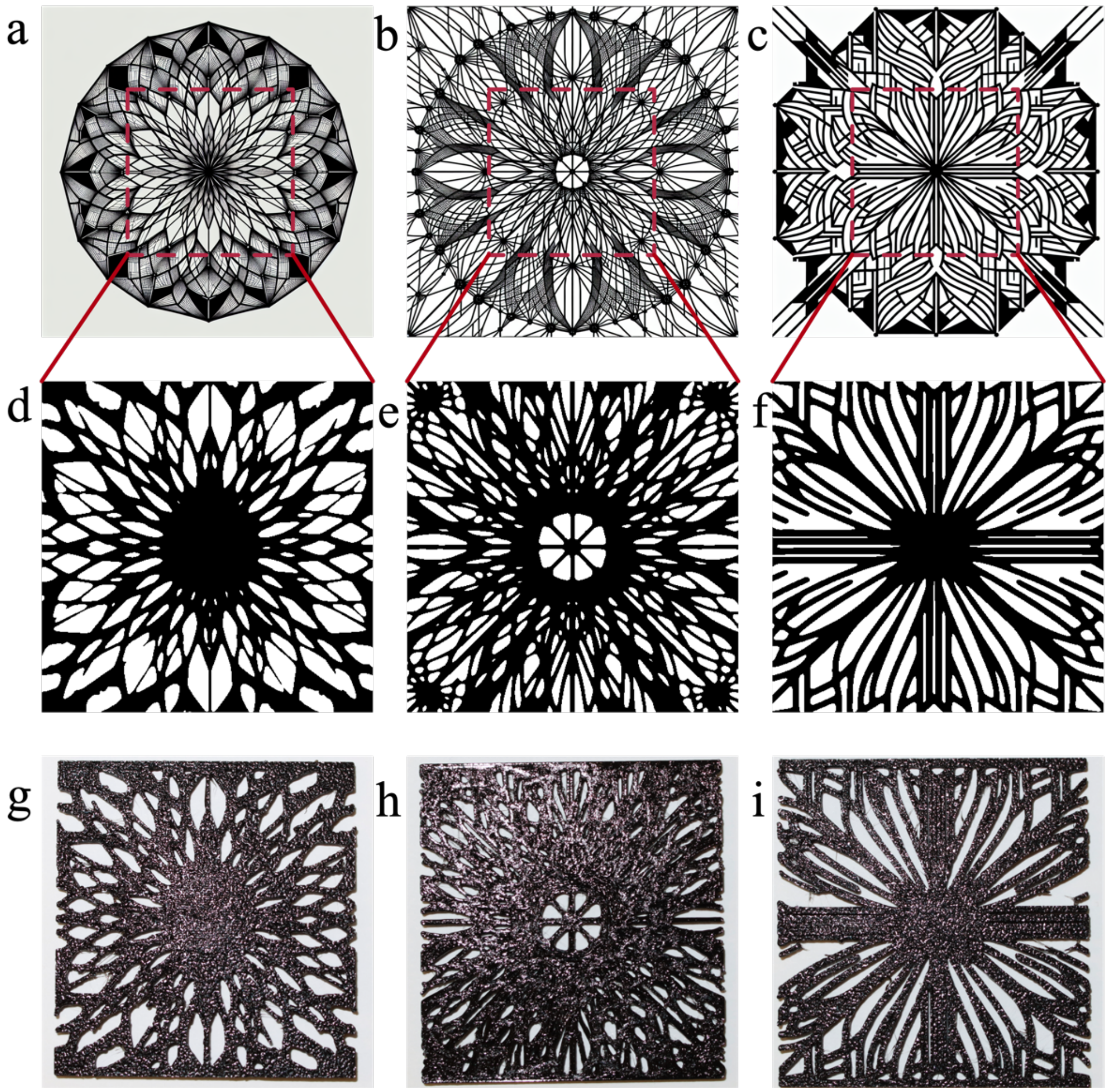

**Figure 12: The demonstration of CrossMatAgent in 3D-print.** (a-c) The three generations from the CrossMatAgent-based MAS show a highly symmetrical pattern and are also very sharp and clear. (d-f) Following the processes of center cropping, binarization, and noise removal, the resultant pattern is found to be identical when compared to the original generation. (g-i) Despite the differences in resolution between the image and the printed model, the generations demonstrate a high degree of manufacturability.

## 4. Conclusions and Future Directions

In this work, we introduced CrossMatAgent, a novel multi-agent framework built upon large language models (LLMs) and multimodal generative models, to streamline and advance the metamaterial design process. By harnessing the synergy of GPT-4o (text) and Dall-E 3 (image) within a carefully constructed agent hierarchy, this approach facilitates the automated collection and augmentation of data, while capturing crucial geometric and morphological information from a diverse range of input sources such as SEM images, 2D/3D rendered models, and photographic representations of metamaterials. A key advancement of this framework lies in the fine-tuning of Stable Diffusion XL (SDXL), guided by a CLIP-based alignment mechanism, to ensure high-fidelity image generation with a deep understanding of geometry—addressing the limitations of text-based LLMs in representing complex metamaterial structures. Our SHAP analyses further demonstrate that the refined CLIP model can robustly identify salient regions and relate them to specific textual descriptions, ensuring a more interpretable connection between geometry, morphology, and functional intent.

By integrating the fine-tuned SDXL model into an agentic architecture, CrossMatAgent becomes capable of producing designs that transition readily into computational simulation and even 3D printing. The incorporation of reflection, iterative feedback loops, and role-specific specialization across Describer, Architect, Builder, and Supervisor agents ensures that generated patterns remain faithful to the mechanical or bio-inspired design principles set by the user. Our simulation results confirm the feasibility of this approach, demonstrating that models derived from CrossMatAgent are meshable, path-connected, and exhibit expected stress distributions under tension, compression, and shear. Notably, the framework also exhibits robust extendibility and few-shot learning capabilities through the adoption of LoRA fine-tuning, showcasing the adaptability of CrossMatAgent to new design styles with minimal additional training data.

While the present study successfully establishes a unified, scalable pathway to metamaterial design grounded in LLMs, several directions remain open for exploration. First, broader data diversity is essential to further enhance CrossMatAgent's coverage and performance. Despite the multi-agent approach's capacity to consolidate information, the current dataset remains relatively modest. Future efforts will focus on systematically integrating larger and more varied collections of SEM images, medical imaging data, and domain-specific morphological structures. Second, incorporating deeper physics- and mechanics-based constraints into the design loop may further improve the practicality of generated samples. For instance, coupling CrossMatAgent with advanced finite element analysis or topology optimization tools could ensure that generated structures not only appear geometrically consistent but also satisfy specific performance criteria. By advancing these directions, CrossMatAgent has the potential to significantly impact the field of metamaterial design, offering a scalable, AI-driven approach that bridges the gap between conceptualization and realization in computational materials science.

## Data Availability Statement

The original contributions presented in the study are included in the article/supplemental material. Supplementary material associated with this article can be found in the online version. Further inquiries can be directed to the corresponding authors. Data is available at https://huggingface.co/datasets/Bio-DMX/CrossMatAgent_MAS_Dataset. The code is available at https://github.com/BioDMX-UGA/CrossMatAgent.git.

## Conflict of Interest

The authors declare that the research was conducted in the absence of any commercial or financial relationships that could be construed as a potential conflict of interest.

## Authorship Contribution Statement

Conceptualization: JT, XW
Methodology: JT, MS, DP, XW
Investigation: JT, MS, DP, KS, XZ, XW
Visualization: JT, MS, JH, AR, XC
Supervision: KS, XW
Writing—original draft: JT, XW

# Supplementary Materials for

## A Multi-Agent Framework Integrating Large Language Models and Generative AI for Accelerated Metamaterial Design

Jie Tian *et al.*

*Corresponding author. Email: kenan.song@uga.edu; xqwang@uga.edu

**This PDF file includes:**

Tables S1 to S4

| Agents | System Prompt |
|---|---|
| **Describer** | Given an image. Analyze the patterns inside.<br>1. Describe the pattern in a general way.<br>2. Describe the pattern as detail as possible from the perspective of structure. If repeat units appears in the pattern, describe the repeat unit.<br>3. Analysze the pattern in a perspective of geometry, seperate the pattern into different geometric parts. If repeat units appears in the pattern, describe the repeat unit.<br>4. Based on the geometric analysis, describe the morphological transformation under tension and compression. |
| **Architect** | You are a professional in prompt engineering and mechanics. Based on the given information, analyze the potential structure of the pattern that satisfies the provided description in both geometric and morphological ways. Describe the two-dimensional pattern. The description should focus on the geometry of the structure. Detail the potential structure according to the description of the shape, size of parts, how the parts are connected, and the location of each part. Keep in mind to ensure the connectivity of the parts. |
| **Builder** | You are a professional in prompt engineering and can write prompts for DALL-E to generate images. These prompts are tailored to create distinct structures of 2D patterns that showcase clear mathematical or structural designs, with an emphasis on maintaining connectivity within the patterns. It specializes in producing patterns that are strictly two-dimensional and ensures each design maintains continuous connectivity, important for applications like tiling and tessellation. This system analyzes key geometric and mathematical elements from a provided image or concept and then generates a new pattern, ensuring it features clean shapes, sharp contrasts, and maintains connectivity throughout the design. The patterns resemble scientific diagrams or theoretical models, exclusively in 2D, and are ideal for replication where continuous connectivity is essential. This system excels in producing intricate, mathematically inspired, visually striking, clearly defined, and consistently interconnected patterns. Use black as the solid color and white as the background color.<br>The viewpoint is 2D, without any 3D simulation or effects.<br>Then draw the pattern based on the prompt. Also, provide the revised prompt for DALL-E.<br>Output format:`## Image\n![image](url)\n## Revised Prompt\nrevised prompt` |
| **Supervisor** | Based on the pattern's description, structure, and the generated output, review the pattern. Review the entire process; check both the description and the architecture, as well as the generated pattern. Assess whether the generated pattern aligns with the description and architecture. If the pattern does not align, provide feedback on how to enhance it. If the pattern is consistent with the description and architecture, offer feedback on its quality and suggest improvements. Also, evaluate the description and architecture, indicating whether they are effective, and provide improvement suggestions. If everything is satisfactory, go to FINISH, and return feedback as None. |

**Table S1:**
System prompts for all agents in multi-agent system

| | |
|---|---|
| **Input** | |
| **Describer** | [AIMessage(content="General_Description='The image displays a highly intricate and dense pattern of overlapping thin fibers or lines. The fibers form a chaotic and random mesh-like structure, giving an organic or web-like appearance.' Structure_Analysis='The structure consists of numerous thin, elongated, and irregularly oriented fibers. These fibers intersect at various angles, creating a dense, interwoven network. There is no clear periodicity or uniformity in the arrangement, and the density of fibers varies slightly throughout the image.' Geometry_Analysis='From a geometrical perspective, the pattern is composed of elongated linear elements (fibers) that intersect and overlap. The fibers exhibit irregular lengths and orientations, forming acute and obtuse angles at intersections. The overall pattern lacks symmetry and follows a stochastic (random) distribution.' Morphological_Transformation='Under tension, the fibers may align more in a particular direction, reducing the randomness and forming a more stretched and less entangled pattern. Under compression, the fibers might buckle or form more folds, increasing the density and entanglement of the network.'", name='Describer')] |
| **Architect** | {'Architect': {'message': [AIMessage(content="The 2-dimensional pattern described can be understood as a non-periodic, web-like structure formed by randomly distributed elongated fibers. Here's a detailed breakdown:\n\n### 1. **Description of the Shape**:\n - **Fibers/Lines**: The primary elements of the structure are thin, elongated, linear components that resemble fibers or threads.\n - **Intersections**: The fibers intersect at irregular angles, forming acute and obtuse connections.\n - **Network Appearance**: The overall structure resembles a chaotic mesh or web pattern, with no consistent directionality or periodic repetition.\n\n### 2. **Size of Parts**:\n - **Fibers**: The fibers vary in length, but each is relatively thin compared to the overall size of the pattern.\n - **Thickness**: The width of each fiber is uniform and very small, contributing to the delicate appearance of the mesh.\n - **Pattern Scale**: The entire pattern can span a large area, with the density of fibers increasing or decreasing locally.\n\n### 3. **Connectivity of Parts**:\n - **Intersection Points**: Fibers connect to one another at intersection points, where they cross each other's paths.\n - **Overlap**: At these intersections, fibers maintain their thin and linear shape; there is no merging or thickening of the material.\n - **Random Distribution**: The fibers do not follow a fixed grid or alignment but are instead placed in a stochastic manner, ensuring a unique connectivity layout.\n\n### 4. **Location of Each Part**:\n - **Fibers**: Distributed randomly across the plane, with varying orientations and lengths.\n - **Intersections**: Scattered throughout the pattern, with some regions showing higher density (more intersections) and others being more sparse.\n - **Clusters**: Certain areas may exhibit clustering of fibers, where multiple fibers converge or overlap in a confined zone, creating a denser mesh-like region.\n\n### Summary:\nThe pattern is an irregular, organic, web-like 2D structure made up of thin, elongated fibers. The fibers are randomly oriented and intersect at various angles, forming a dense, interconnected network. The lack of periodicity and uniformity suggests a natural or stochastic arrangement. The fibers maintain a consistent thickness and are connected at discrete points without merging. This |
| | description ensures that the connectivity and geometric properties of the structure are preserved.", name='Architect')], 'name': 'Architect', 'send_to': None}} |
| **Builder** | tool='dallewraper' tool_input={'prompt': 'A 2D intricate interconnected pattern resembling a scientific diagram or theoretical model. The design consists of thin, elongated black lines (fibers) on a white background forming a dense, non-periodic mesh-like network. The fibers intersect at irregular angles, creating acute and obtuse connections, maintaining continuous connectivity throughout the design. The structure appears chaotic yet meticulously interconnected, emphasizing sharp contrast and clean geometric shapes. Exclusively in black and white, no 3D effects or shading. Ideal for tiling or tessellation, resembling a mathematical or scientific conceptual pattern.', 'model': 'dall-e-3', 'size': '1024x1024', 'response_format': 'url'} log="\nInvoking: `dallewraper` with `{'prompt': 'A 2D intricate interconnected pattern resembling a scientific diagram or theoretical model. The design consists of thin, elongated black lines (fibers) on a white background forming a dense, non-periodic mesh-like network. The fibers intersect at irregular angles, creating acute and obtuse connections, maintaining continuous connectivity throughout the design. The structure appears chaotic yet meticulously interconnected, emphasizing sharp contrast and clean geometric shapes. Exclusively in black and white, no 3D effects or shading. Ideal for tiling or tessellation, resembling a mathematical or scientific conceptual pattern.', 'model': 'dall-e-3', 'size': '1024x1024', 'response_format': 'url'}`\n\n\n" message_log=[AIMessageChunk(content='', additional_kwargs={'function_call': {'arguments': '{"prompt":"A 2D intricate interconnected pattern resembling a scientific diagram or theoretical model. The design consists of thin, elongated black lines (fibers) on a white background forming a dense, non-periodic mesh-like network. The fibers intersect at irregular angles, creating acute and obtuse connections, maintaining continuous connectivity throughout the design. The structure appears chaotic yet meticulously interconnected, emphasizing sharp contrast and clean geometric shapes. Exclusively in black and white, no 3D effects or shading. Ideal for tiling or tessellation, resembling a mathematical or scientific conceptual pattern.","model":"dall-e-3","size":"1024x1024","response_format":"url"}', 'name': 'dallewraper'}}, response_metadata={'finish_reason': 'function_call'}, id='run-a0d3dd6c-1940-4bc9-b78b-fb44025f0f62')] |
| | 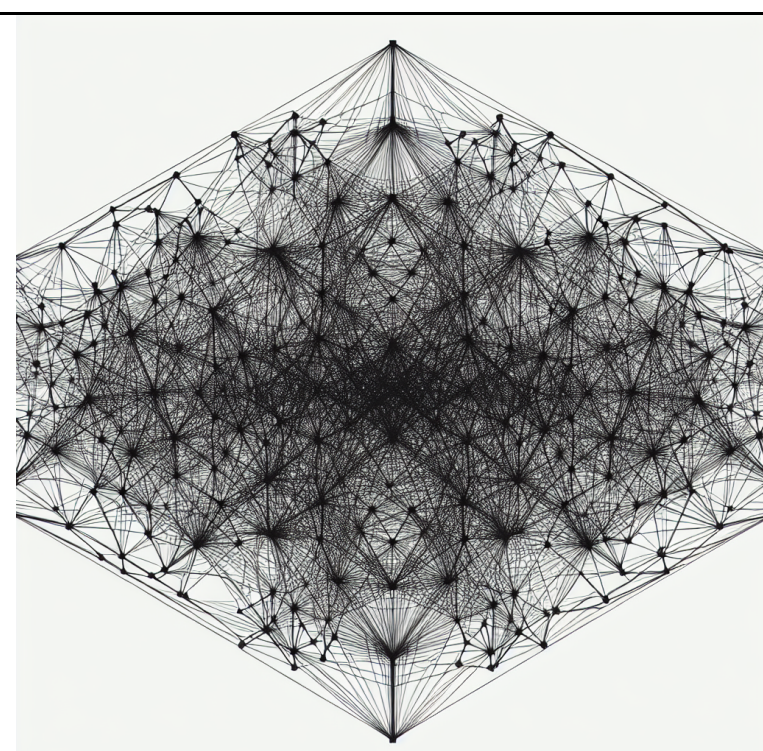 |
| **Supervisor** | {'Reviewer': {'message': [AIMessage(content="The 2-dimensional pattern described can be understood as a non-periodic, web-like structure formed by randomly distributed elongated fibers. Here's a detailed breakdown:\n\n### 1. **Description of the Shape**:\n   - **Fibers/Lines**: The primary elements of the structure are thin, elongated, linear components that resemble fibers or threads.\n   - **Intersections**: The fibers intersect at irregular angles, forming acute and obtuse connections.\n   - **Network Appearance**: The overall structure resembles a chaotic mesh or web pattern, with no consistent directionality or periodic repetition.\n\n### 2. **Size of |
| | Parts**:\n   - **Fibers**: The fibers vary in length, but each is relatively thin compared to the overall size of the pattern.\n   - **Thickness**: The width of each fiber is uniform and very small, contributing to the delicate appearance of the mesh.\n   - **Pattern Scale**: The entire pattern can span a large area, with the density of fibers increasing or decreasing locally.\n\n### 3. **Connectivity of Parts**:\n   - **Intersection Points**: Fibers connect to one another at intersection points, where they cross each other's paths.\n   - **Overlap**: At these intersections, fibers maintain their thin and linear shape; there is no merging or thickening of the material.\n   - **Random Distribution**: The fibers do not follow a fixed grid or alignment but are instead placed in a stochastic manner, ensuring a unique connectivity layout.\n\n### 4. **Location of Each Part**:\n   - **Fibers**: Distributed randomly across the plane, with varying orientations and lengths.\n   - **Intersections**: Scattered throughout the pattern, with some regions showing higher density (more intersections) and others being more sparse.\n   - **Clusters**: Certain areas may exhibit clustering of fibers, where multiple fibers converge or overlap in a confined zone, creating a denser mesh-like region.\n\n### Summary:\nThe pattern is an irregular, organic, web-like 2D structure made up of thin, elongated fibers. The fibers are randomly oriented and intersect at various angles, forming a dense, interconnected network. The lack of periodicity and uniformity suggests a natural or stochastic arrangement. The fibers maintain a consistent thickness and are connected at discrete points without merging. This description ensures that the connectivity and geometric properties of the structure are preserved.", name='Architect'), HumanMessage(content=[{'type': 'image_url', 'image_url': {'url': 'https://oaidalleapiprodscus.blob.core.windows.net/private/org-aixs1fqEEt7RuFxXUGghZppC/user-ezot9SEdVtJQTp35ynXFKSrs/img-52hVe6cnIeYmrzukFKZTzkRA.png?st=2025-01-14T20%3A58%3A53Z&se=2025-01-14T22%3A58%3A53Z&sp=r&sv=2024-08-04&sr=b&rscd=inline&rsct=image/png&skoid=d505667d-d6c1-4a0a-bac7-5c84a87759f8&sktid=a48cca56-e6da-484e-a814-9c849652bcb3&skt=2025-01-14T00%3A40%3A30Z&ske=2025-01-15T00%3A40%3A30Z&sks=b&skv=2024-08-04&sig=a2rZTFFtiV1WXV6PGc91P4v7J3M%2B5jOs8nCiN6gQ5Jg%3D'}}], additional_kwargs={'revised_prompt': "An intricate 2D interconnected pattern that recalls a scientific diagram or theoretical model. The design includes thin, elongated black lines, akin to fibers, against a pure white background. These lines form a dense, non-periodic mesh-like network. They intersect at both acute and obtuse angles, ensuring that a sense of continuous connectivity pervades throughout the design. The overall structure lands somewhere between chaotic and methodically interconnected, and it showcases an attention-grabbing contrast and clean geometric shapes. The image uses solely black and white and has no 3D effects or shading. It's perfectly suited for tiling or tessellation, and it's reminiscent of a mathematical or scientific conceptual pattern."}, name='Builder'), AIMessage(content='The generated pattern, while intricate and interconnected, deviates significantly from the initial random fiber mesh structure described and architected. Instead of a chaotic web-like pattern, the result appears to be a structured network with geometric symmetry and node-based connectivity. Consider revising the pattern generation process to create a more randomized, organic appearance with overlapping fibers intersecting at various angles and no clear symmetry or nodal arrangement. This will better align with the original description and architectural intent.', name='Reviewer')], 'name': 'Reviewer', 'send_to': 'Builder'}} |
| **Builder** | tool='dallewraper' tool_input={'prompt': 'A 2D intricate web-like pattern with randomly distributed elongated fibers and no clear symmetry. The fibers are thin and black, intersecting at irregular angles to form a continuous interconnected chaotic mesh. The background is solid white. The entire pattern resembles an organic, stochastic layout with overlapping fibers, avoiding structured geometrical symmetry.', 'size': '1024x1024', 'response_format': 'url'} log="\nInvoking: |
| | `dallewraper` with `{'prompt': 'A 2D intricate web-like pattern with randomly distributed elongated fibers and no clear symmetry. The fibers are thin and black, intersecting at irregular angles to form a continuous interconnected chaotic mesh. The background is solid white. The entire pattern resembles an organic, stochastic layout with overlapping fibers, avoiding structured geometrical symmetry.', 'size': '1024x1024', 'response_format': 'url'}`\n\n\n" message_log=[AIMessageChunk(content=", additional_kwargs={'function_call': {'arguments': '{"prompt":"A 2D intricate web-like pattern with randomly distributed elongated fibers and no clear symmetry. The fibers are thin and black, intersecting at irregular angles to form a continuous interconnected chaotic mesh. The background is solid white. The entire pattern resembles an organic, stochastic layout with overlapping fibers, avoiding structured geometrical symmetry.","size":"1024x1024","response_format":"url"}', 'name': 'dallewraper'}}, response_metadata={'finish_reason': 'function_call'}, id='run-3754f9fa-7eb3-48d7-9658-d7286253aa98')] |
| | 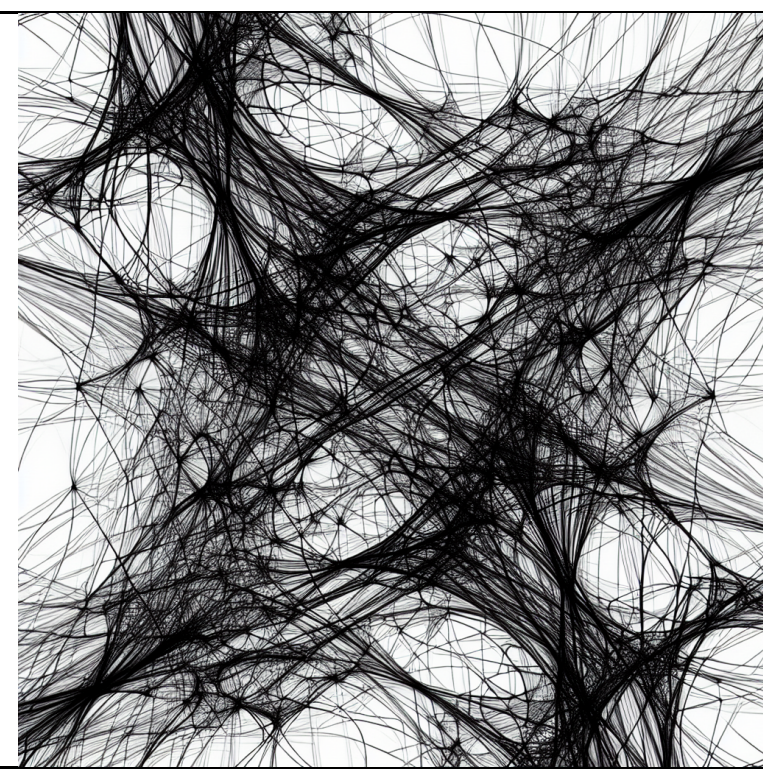 |

**Table. S2**

The Chat History Example of MAS.

| Image | Prompt for Dall-E 3 | Prompt for SDXL |
|---|---|---|
| 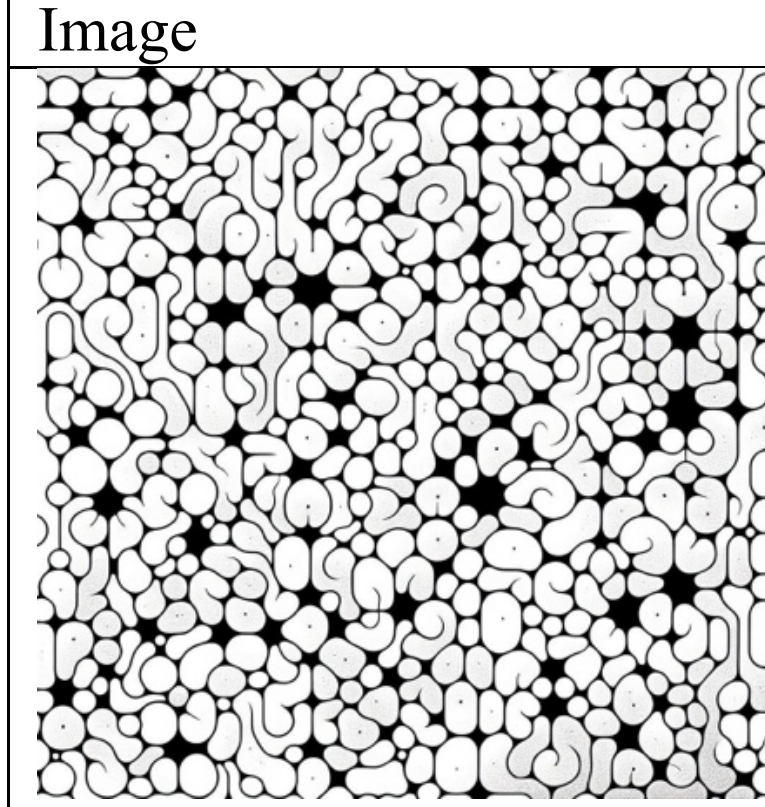 | Create a 2D pattern with units that resemble irregular, organic-shaped blobs or cells, varied in size and form to produce a non-repetitive and random arrangement. Spaces between these units are outlined by thin, dark lines that ensure connection throughout the design. The overall configuration should appear dense with minimal gaps, as though it's a meticulously arranged puzzle of irregular polygons featuring curved edges. Use black as the dominant color for the shapes in contrast to a white background. The | Create, 2D pattern, irregular organic-shaped blobs or cells, varied size and form, non-repetitive random arrangement, thin dark lines, connection throughout, dense configuration, minimal gaps, black shapes, white background, plainly 2D, no 3D effect |
| | composition should be plainly 2D, without any simulation of a 3D effect. | |
| 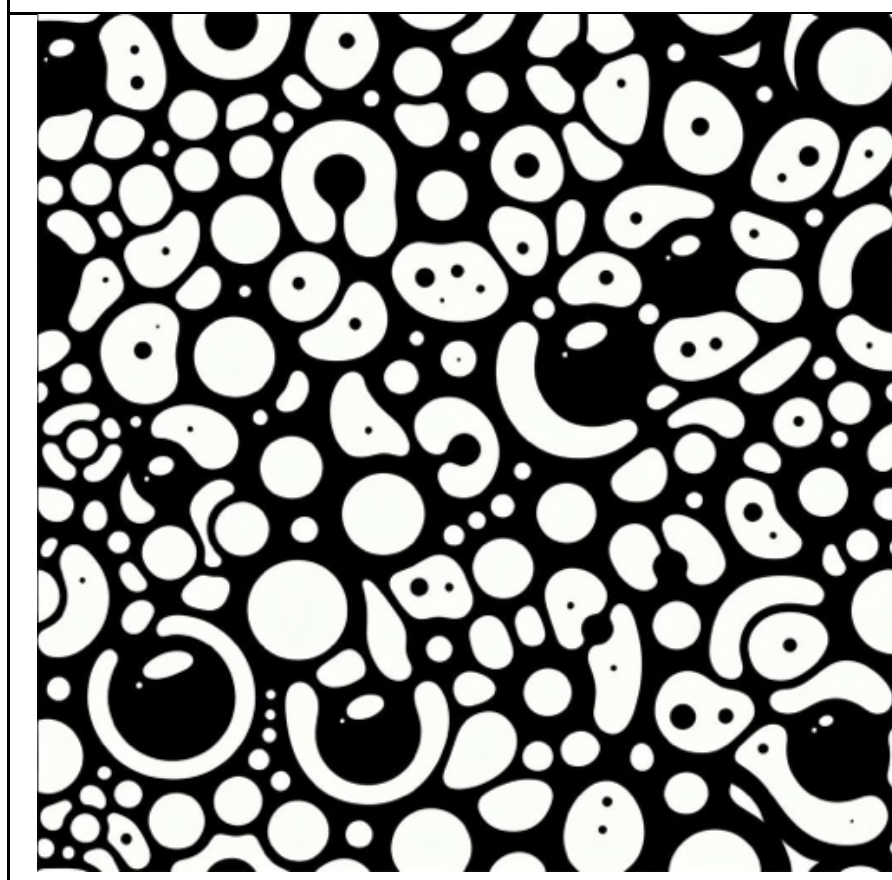 | Create an image depicting a 2D pattern consisting of irregular, organic, cell-like shapes of various sizes with rounded edges. The shapes should be interconnected by thick lines, providing a sense of a cohesive network. The pattern should embrace a random orientation, no symmetry or repetitive elements. Colors should be bold and stark, using black for the cell-like shapes and lines, and white as the background color. | 2D pattern, irregular, organic, cell-like shapes, various sizes, rounded edges, interconnected, thick lines, cohesive network, random orientation, no symmetry, no repetitive elements, bold colors, black shapes, black lines, white background |
| 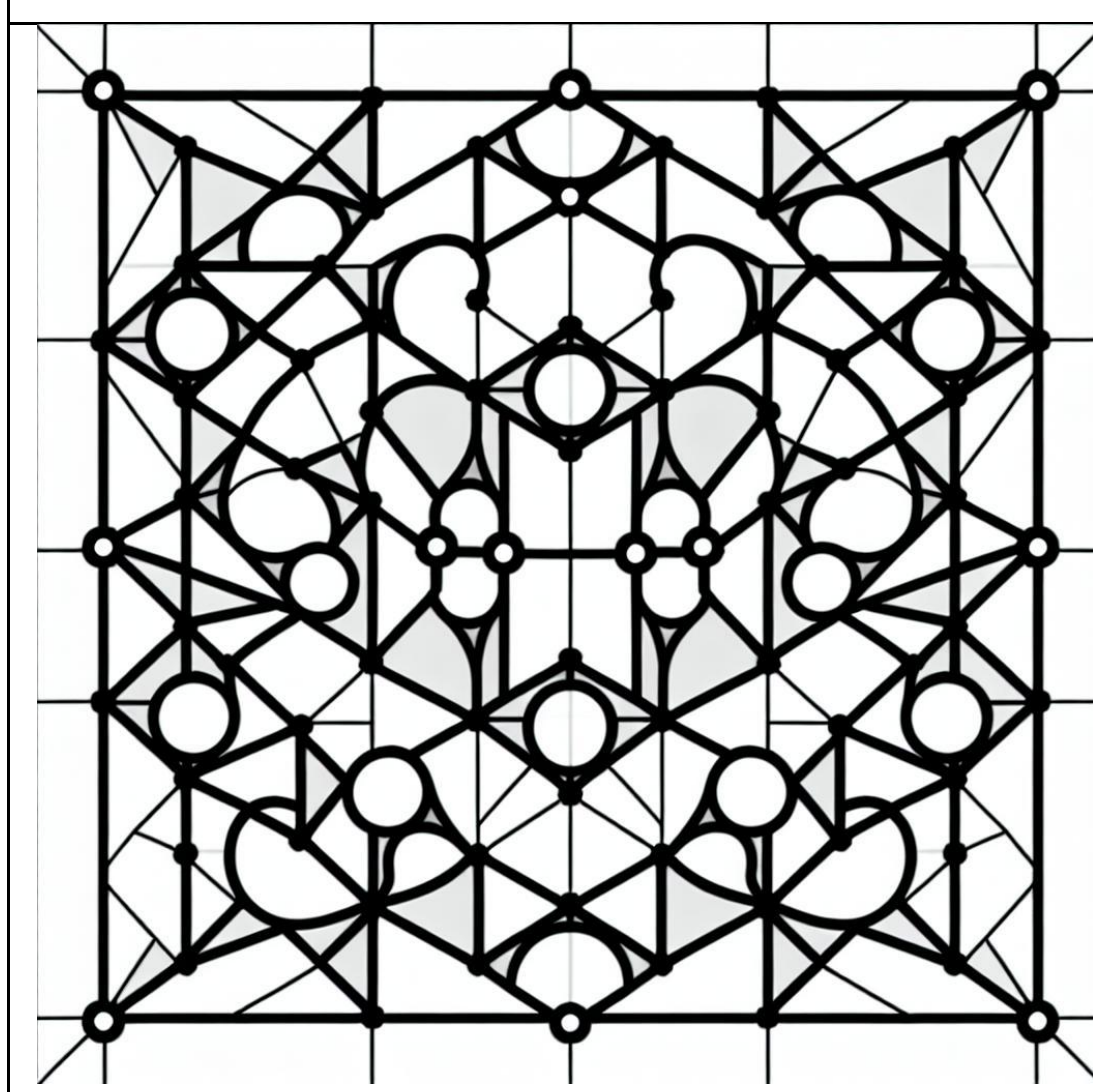 | Draw a black and white, 2D geometric pattern consisting of interconnected, tessellated triangular units. Each triangle should have slightly curved sides and a circular element at its center. Arrange the units in such a way that each one is rotated 90 degrees relative to its adjacent triangle, creating a continuously connected and interlocking design. The overall pattern should resemble a scientific diagram with clean shapes, sharp contrasts, and clear connections throughout. Fill the background with white color while the geometric shapes are in black. | black and white, 2D geometric pattern, interconnected tessellated triangles, slightly curved sides, circular center, each rotated 90 degrees, continuously connected interlocking design, scientific diagram-like, clean shapes, sharp contrasts, clear connections, white background, black geometric shapes |

**Table S3**

Data Examples of Dataset, total dataset is available at https://huggingface.co/datasets/Bio-DMX/CrossMatAgent_MAS_Dataset

| **Prompt for ChatGPT** | **Generated Prompt for Dall-E** |
|---|---|
| Based on the pattern of given image, generate a new pattern of 2D metamaterial. | Create a 2D geometrical pattern containing the elements of interlocking isosceles triangles and inscribed circular arcs. The triangles should be arranged in a tessellated pattern where the apex of each triangle connects with the bases of the surrounding triangles. Within each triangle, a circular arc should be drawn with its endpoints touching the midpoints of the triangle's base. The pattern should be symmetrical and exhibit continuous connectivity for a harmonious aesthetic. Use black as the main solid color for the geometrical shapes, contrasting against a stark white background. |
| Based on the pattern of given image. Analyze the structure of the pattern and generate a new pattern of 2D metamaterial have similar pattern. | Create a 2D geometric pattern that features interconnected units. Each unit is composed of an isosceles triangle where the two equal sides are longer than the base. Within each triangle is nestled a perfectly fitting circular arc, whose endpoints touch each side of the triangle. The units follow a specific arrangement, rotated and tessellated to ensure seamless interlocking and continuous connectivity. The pattern's symmetry is twofold: translational and rotational, creating a visual harmony resembling scientific diagrams. Sharp contrast and clean lines feature prominently. The geometric pattern should be presented in black, contrasted against a white background. The view angle is strictly orthogonal, reinforcing the 2-dimensional design. |
| Based on the pattern of given image, Analyze the structure of the pattern from the perspective of morphological and mechanical properties, then generate a new pattern of 2D. | Create a black and white 2D geometric pattern. The pattern should be comprised of interlocking isosceles triangles and inscribed circular arcs. The triangles should be tessellated, creating a seamless design. Each triangle should connect at its peak to the bases of neighboring triangles, with each circular arc fitting perfectly within the boundaries of the triangle, connecting at the midpoints of the triangle's base. The pattern should repeat with symmetry in both the horizontal and |
| | vertical directions, maintaining high structural integrity. |

**Table S4**

Prompt for generation using the chatbox.